\documentclass[12pt]{iopart}

\usepackage{amsthm}
\usepackage{amssymb}
\usepackage{iopams}  
\usepackage{epsfig}
\usepackage{verbatim}
\usepackage{hyperref}
\usepackage{harvard}
\usepackage{url}
\begin{document}

\title[Increasing LIGO sensitivity by feedforward subtraction]{Increasing LIGO sensitivity by feedforward subtraction of auxiliary length control noise}

\author{Grant David Meadors$^1$, Keita Kawabe$^2$, Keith Riles$^1$}

\address{$^1$ Department of Physics, University of Michigan, 450 Church Street, Ann Arbor, MI~48109, US}
\address{$^2$ LIGO Hanford Observatory, PO Box 159, Richland, WA 99352-0159, US}
\eads{\mailto{gmeadors@umich.edu}, \mailto{kawabe\_k@ligo-wa.caltech.edu}, \mailto{kriles@umich.edu}}
\begin{abstract}
LIGO, the Laser Interferometer Gravitational-wave Observatory, has been designed and constructed to measure gravitational wave strain via differential arm length. The LIGO 4-km Michelson arms with Fabry-Perot cavities have auxiliary length control servos for suppressing Michelson motion of the beam-splitter and arm cavity input mirrors, which degrades interferometer sensitivity. We demonstrate how a post-facto pipeline (AMPS) improves a data sample from LIGO Science Run~6 with feedforward subtraction. Dividing data into 1024-second windows, we numerically fit filter functions representing the frequency-domain transfer functions from Michelson length channels into the gravitational-wave strain data channel for each window, then subtract the filtered Michelson channel noise (witness) from the strain channel (target). In this paper we describe the algorithm, assess achievable improvements in sensitivity to astrophysical sources, and consider relevance to future interferometry.

\end{abstract}

\maketitle

    \section{Introduction}
    \label{introduction}

Antennae for gravitational wave observations~\cite{Thorne300} require precise understanding of noise sources to attain peak sensitivity. Some of these noises arise from auxiliary degrees of freedom in interferometric antennae. Feedforward control can correct these auxiliary control noises. Cluster computing on archived data makes previous methods of feedforward correction scalable to year-long science runs. Computing can also adjust for the non-stationarity inherent in these noise couplings. This paper describes such a computational method and the improvements it might provide for searches with LIGO (Laser Interferometer Gravitational-wave Observatory).

	As a network with GEO600~\cite{Willke2002,Hild2009} and VIRGO~\cite{Acernese2005}, Enhanced LIGO~\cite{LIGOFirst2004,Fricke2009} produced data during LIGO Science Run 6 (S6) that was the most sensitive yet taken in the search for gravitational waves of astrophysical origin reaching the Earth: in this paper, we further enhance LIGO sensitivity via post-run software corrections. Radio astronomy of pulsar systems such as PSR 1913+16~\cite{HulseTaylor1975,Weisberg2010} provides indirect evidence for gravitational radiation, and direct detections could elucidate the structure of neutron stars~\cite{Lindblom1995,AbbottPulsar2006} and illuminate black holes~\cite{Sathyaprakash2009}, supernovae~\cite{Chandrasekhar1969,Ott2009}, cosmology~\cite{Grishchuk1974}, and related tests of the strong-field validity of general relativity~\cite{Riles2013}. This potential motivates new observatories, such as KAGRA~\cite{Kuroda2010}, and improvements to existing observatories. 

LIGO infers gravitational-wave strain $h(t)$ at each of its two observatories [Hanford, Washington and Livingston, Louisiana] from the length difference between 4-km Michelson interferometer arms~\cite{Saulson} using a calibration response function~\cite{LIGOCal2010}. Each arm contains a Fabry-Perot resonant cavity locked using the Pound-Drever-Hall technique~\cite{Drever1983,Black2001}, comprised of an input test mass, near the Michelson beam-splitter, and an end test mass. A power-recycling mirror sits between the laser and the beam-splitter. These six core optics form coupled optical cavities with four length degrees of freedom, each of which is servoed to maintain optical resonance by minimizing motion (see Section~\ref{motive_math}). The effective change in the differential arm length $L_-$ (colloquially DARM) caused by gravitational waves is encoded in the intensity of the light reaching the anti-symmetric port of the Michelson interferometer and is read out by DC homodyne~\cite{Fricke2009}. Auxiliary length control for the beam-splitter and input mirrors is becoming more complex, as in Advanced LIGO, which will add a signal recycling cavity. This paper describes post-facto software improvements of detector noise using adaptive feedforward subtraction in a pipeline called Auxiliary MICH-PRC Subtraction (AMPS)~\cite{MatappsRepository}: these improvements refine LIGO's gravitational-wave sensitivity to astrophysical sources.

AMPS improves LIGO S6 data (2009 July 07 to 2010 October 20), as this paper will show. S6 gravitational wave strain (target) is corrected based on auxiliary length noise measurements (witness). Enhanced LIGO generated the S6 data with high laser power and DC readout to prepare for Advanced LIGO. The motion of the beam-splitter and input mirrors of the Fabry-Perot cavities is known~\cite{AdhikariThesis,BallmerThesis} to cause cross-talk in the gravitational wave strain channel, which compounds a noise floor fundamentally limited by seismic, thermal suspension, and laser shot noise. Observed S6 cross-talk included differential Michelson (MICH) as well as power-recycling cavity length (PRC). The DARM readout, as explained in Section~\ref{motive_math}, is intrinsically sensitive to MICH divided by a factor of arm cavity gain, Equation~\ref{rcfactor}. (Theoretically, physical $h(t)$ is imprinted in MICH, but the cavity gain and relative smallness of the Michelson cavity make the effect about five orders of magnitude smaller than in DARM, so it is ignored). Methods~\cite{KisselThesis} to tune real-time feedforward filters for LIGO servo cross-talk are our starting point, but we seek to automate and improve retuning.

Post-facto, adaptive feedforward simplifies cross-talk subtraction. AMPS uses Matlab 2012a~\cite{Matlab2012a}. The witness-to-target transfer function is estimated in discrete time windows of 1024 seconds and fit to a zero-pole-gain filter with Vectfit~\cite{Deschrijver2008,Gustavsen1999,Gustavsen2006}. Safeguards ensure a statistically significant fit that does not further degrade the target signal. Noise from the witnesses passes through respective filters, then is subtracted from the strain target channel. The correction lowers the noise floor, benefitting any gravitational-wave searches using this data.

    \section{Description of the feedforward method}
    \label{motive_math}

Gravitational-wave antennae around the world share features and form a collaborative network. Amongst kilometer-scale Michelson interferometers, GEO600 in Hannover, Germany uses folded arms with both power- and signal-recycling, LIGO, and VIRGO use Fabry-Perot cavities coupled with power- (and potentially signal-) recycling cavities. The Japanese interferometer KAGRA, under construction, will have a similar optical layout to LIGO and VIRGO but with cryogenically-cooled mirrors in an underground laboratory. Although nomenclature here pertains to LIGO, the core problem of this paper applies directly to all power-recycled Michelson interferometers with Fabry-Perot arms. It could extend to other instruments with multiple degrees of freedom that obtain a signal from a target contaminated by control noise from auxiliary degrees of freedom, especially when those auxiliaries are controlled using a lower signal-to-noise ratio (SNR) error signal than for the target and when the witnesses are highly independent.

LIGO core optics include the beam-splitter (BS) and power-recycling mirror (PRM), which is situated between the laser and the beam-splitter. The four LIGO mirror test masses (TM) are named by arm (X or Y) and input (I) vs end (E) of the Fabry-Perot cavities. LIGO controls four optical pathlength degrees of freedom~\cite{ReadoutGWA}. DARM is a signal of \textit{differential arm} length, which is calibrated into the primary part of the gravitational strain measurement, \textit{h(t)}. CARM yields \textit{common arm} length, and is controlled with a common mode servo using laser frequency. MICH \textit{Michelson} and PRC \textit{power-recycling cavity} length refer only to input test masses.

        \begin{eqnarray}
        \textup{Strain: } h(t) = \frac{\delta \left(L_{-}(t) \right)}{\langle L_{+}\rangle}, \label{hoftDef}
        \end{eqnarray}
        \begin{eqnarray}
        \textup{Common arm length: } \textup{CARM} \propto \delta(L_{+}) = \frac{\delta(L_y + L_x)}{2}, \label{CARMdef} \\
        \textup{Differential arm length: } \textup{DARM} \propto \delta(L_{-}) = \delta(L_y - L_x), \label{DARMdef} \\
        \textup{Power-recycling cavity length: }\textup{PRC} \propto \delta(l_{+}) = \frac{\delta(l_y + l_x)}{2}, \label{PRCdef} \\
        \textup{(Inner) Michelson length: } \textup{MICH} \propto \delta(l_{-}) = \delta(l_y - l_x), \label{MICHdef}
        \end{eqnarray}

        \begin{eqnarray}
        L_y \equiv z(\textup{ETMY}) - z(\textup{ITMY}), \label{Lydef} \\
        L_x \equiv z(\textup{ETMX}) - z(\textup{ITMX}), \label{Lxdef} \\
        l_y \equiv z(\textup{ITMY}) - z(\textup{RM}), \label{lydef} \\
        l_x \equiv z(\textup{ITMX}) - z(\textup{RM}). \label{lxdef}
        \end{eqnarray}

\begin{figure}
\begin{center}
\includegraphics[height=135mm,width=150mm]{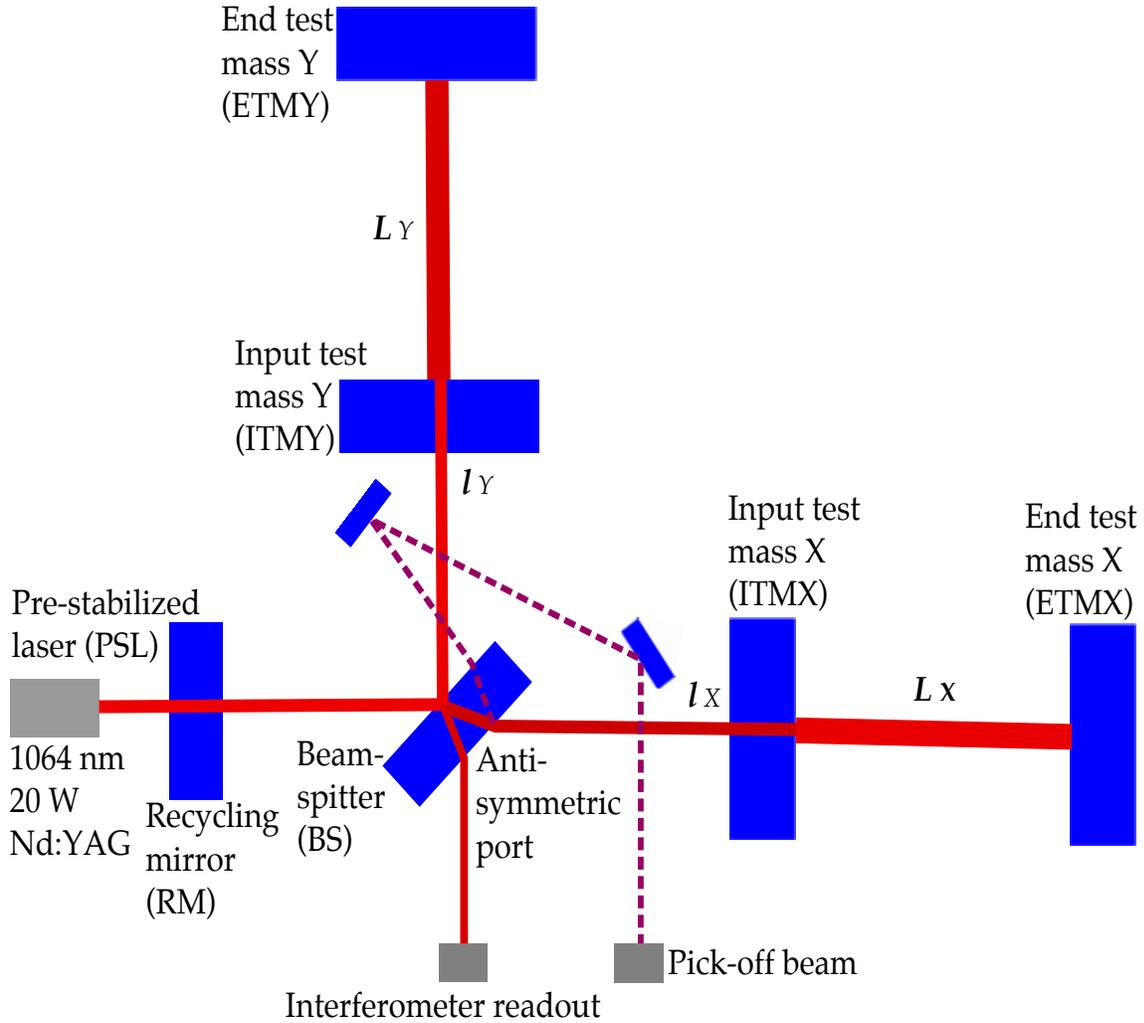}
\caption{Gravitational wave strain $h(t)$ is derived from differential arm motion (DARM), read-out from a photodiode downstream of the antisymmetric port. An internal reflection off an anti-reflective coating, on either the beam-splitter (BS) or an input test mass (ITM), provides the Michelson (MICH) channel. The DARM readout channel predominantly measures the small change in different arm length, $\delta(L_-) \equiv \delta(L_y - L_x)$, while MICH measures that in the Michelson length $\delta(l_-) \equiv \delta(l_y - l_x)$. There is also a small coupling from $\delta(l_-)$ to the DARM channel. To a lesser extent, changes in the length of PRC, which is defined as $\delta(l_+) \equiv \delta(l_y + l_x)/2$ and is measured in quadrature demodulation with respect to the MICH pick-off, also add noise to DARM.}
\label{arms}
\end{center}
\end{figure}

        Average arm length is $\langle L_+\rangle$, about 4 km in LIGO. The distance function $z(\mathcal{X})$ indicates the distance (note that $z(\textup{ITMY})$ is a function of both the ITMY and BS position), along the optical path, from the laser to an optic $\mathcal{X}$. The variation $\delta$ denotes a change with respect to nominal value. DARM length is thus defined as $\delta(L_y - L_x)$ and MICH length as $\delta(l_y - l_x)$. In practice, DARM and MICH are the names given to the channels that predominantly measure those quantities. Unless stated otherwise, the terms DARM, MICH, PRC and CARM will refer to the measured channels, which are related to the lengths through calibration and are cross-contaminated (e.g., DARM = $\delta(L_{-}) + \pi/(2 \mathcal{F}) \delta(l_{-})$, where $\mathcal{F}$ is cavity finesse). The terms will not refer to the ideal physical lengths in Equations~\ref{CARMdef} through~\ref{MICHdef}. 

        As Equation~\ref{DARMdef} and~\ref{MICHdef} imply and Figure~\ref{arms} illustrates, MICH noise ambiguates the physical interpretation of DARM. An arm cavity gain $r_{c}'/r_c \simeq 139/0.990$, where $r_c$ is the arm cavity reflectivity for the LIGO laser carrier frequency and $r_{c}'$ is the derivative of $r_c$ with respect to round trip phase~\cite{ReadoutGWA,BallmerThesis}, amplifies DARM motion for Initial and Enhanced LIGO. Where $\mathcal{F} \simeq 219$, the gain is given by Equation~\ref{rcfactor}:

        \begin{eqnarray}
        \frac{r_{c}'}{r_c} = \frac{2 \mathcal{F}}{\pi} \simeq (139/0.990). \label{rcfactor}
        \end{eqnarray}

        \textit{A priori} MICH noise will leak into measurements of DARM with a transfer function equal to the inverse of Equation~\ref{rcfactor}~\cite{SiggFreq1997}. Coherence measurements confirm this coupling dominates the transfer function, but residuals suggest other effects exist. PRC is also indirectly correlated with DARM. These correlations are physical consequences of the interferometer design. 

In Enhanced LIGO, DARM is measured with a photodiode at the interferometer `dark' antisymmetric port of the beams-splitter. Independent photodiodes for MICH and PRC, used for feedback on their respective auxiliary length control servos, provide the witness channels for canceling cross-talk into DARM. The MICH and PRC photodiodes receive a beam from an internal reflection in the beam-splitter. This beam carries a radio-frequency modulation; one demodulation quadrature provides MICH, the other PRC.

        \subsection{Auxiliary noise coherence at sensitive frequencies}
        \label{aux_noise}

	 Coherence, the Fourier frequency-dependent analog of statistical covariance, quantifies cross-talk. On a scale of 0 (none) to 1 (full), magnitude-squared-coherence (MS-coherence) represents the normalized fraction of power of a frequency bin in the spectrum of one channel that can be found in the same frequency bin in the spectrum of another channel. First, we must define the cross-power of a two time-series. Where $P_{xy}$ is cross-power spectral density, 
%
%
%
we can describe how the coherence at a given frequency $f$ and time $t$~\cite{Boashash1990} is given by Equation~\ref{coherenceDef}. 
\begin{eqnarray}
C_{xy}(f, t) = \sqrt{\frac{\left| P_{xy}(f, t) \right|^2}{P_{xx}(f, t) P_{yy} (f, t)}}\label{coherenceDef}.
\end{eqnarray}



\begin{figure}
\begin{center}
\includegraphics[height=80mm, width=75mm]{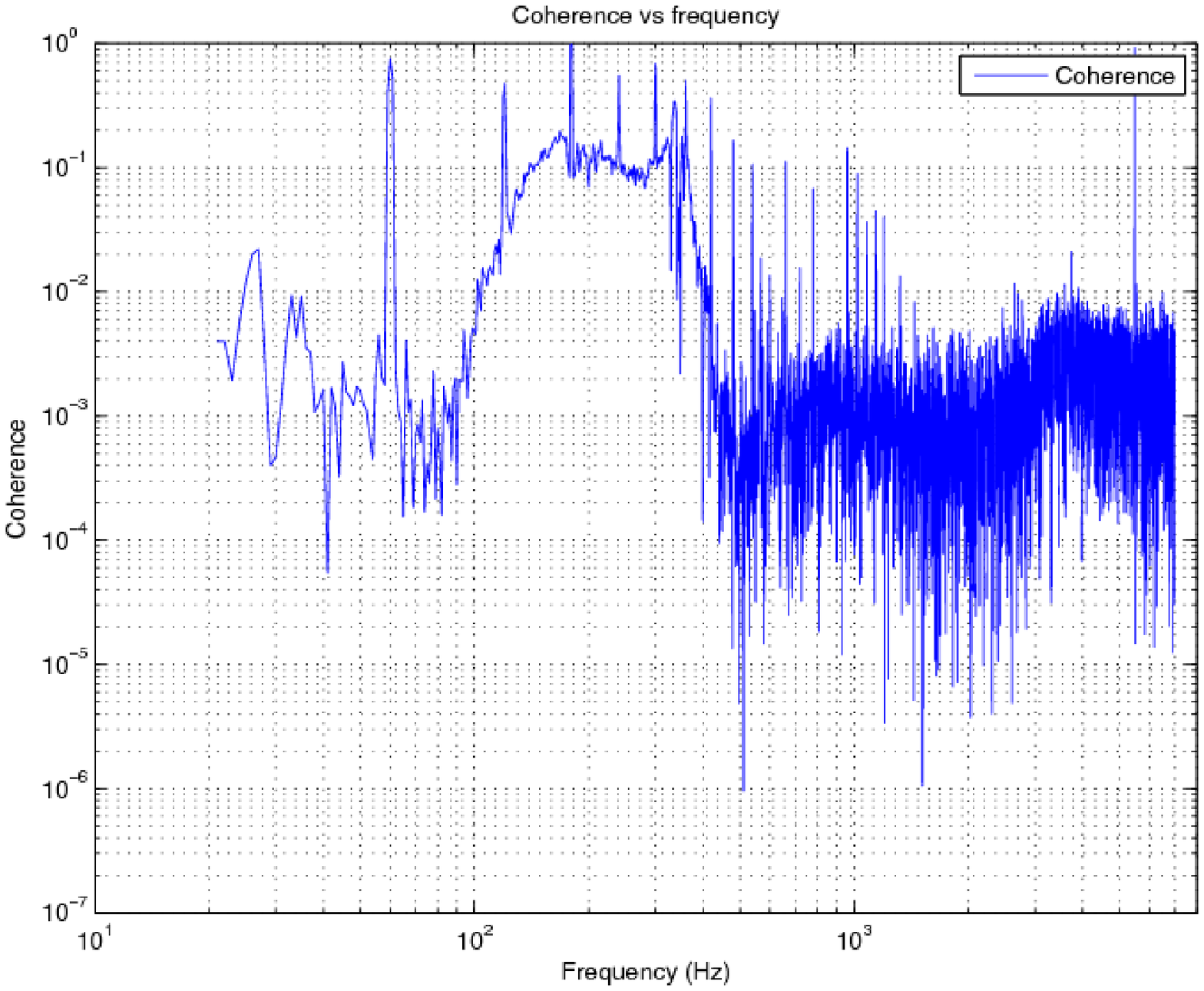}
\includegraphics[height=80mm, width=75mm]{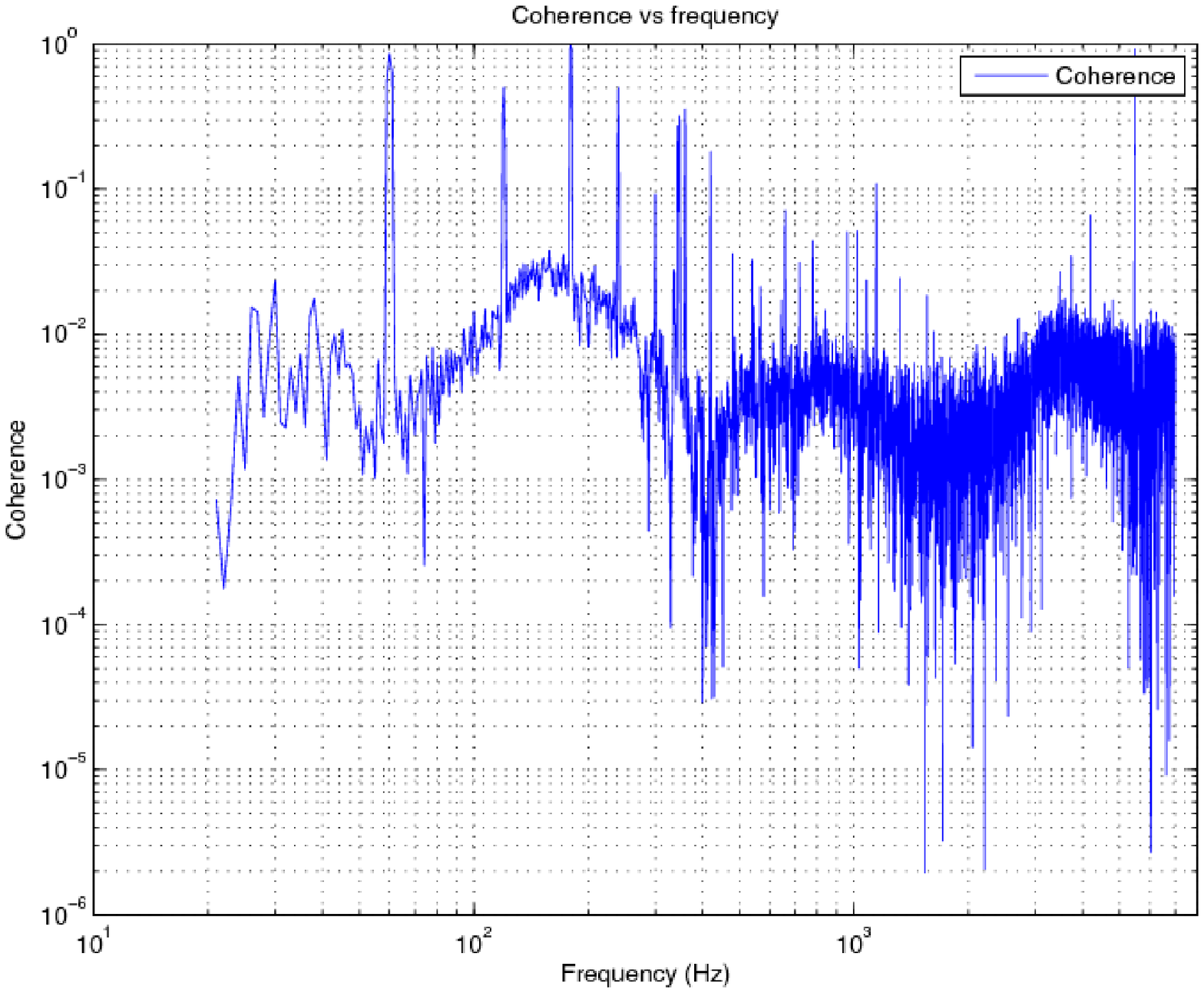}
\caption{Sample coherence measurements between $h(t)$ and auxiliary control channels for LIGO Hanford Observatory, H1: 2010 March 21. MICH-$h(t)$ coherence on left, PRC-$h(t)$ coherence on right. Statistically significant coherence justifies fitting; in frequency bands, about 80 to 400 Hz, where coherence rose above background levels, the transfer function fit was weighted more heavily. Units of coherence spectral density (Hz$^{-1/2}$) vs frequency (Hz).}
\label{coherenceGraph}
\end{center}
\end{figure}

The calibrated strain channel for $h(t)$ (internally, the discrete-time calibrated strain channel for the physical strain $h(t)$ is called `Hoft'), is, with high confidence, coherent with MICH and PRC, as seen in Figure~\ref{coherenceGraph}. MICH-$h(t)$ coherence is sometimes as large as 0.1 in the 100 to 300 Hz band; PRC-$h(t)$ is an order of magnitude lower. Unfortunately, this is the most sensitive band for Initial and Enhanced LIGO.

        \subsection{Estimating filters}
        \label{filter_est}

	Allen, Hua, and Ottewill (AHO)~\cite{AllenHuaOttewill1999} proposed the filtering scheme that this paper employs. Where there is a strong correlation between a signal (target) channel and a noise (witness) channel, the noise can be partially cancelled if a witness-to-target transfer function, convolved with the witness, is applied to the measured target.
	Equations~\ref{gcf} through~\ref{hatsf} capture this method. It is analogous to frequency-domain principal component analysis (PCA) using Gram-Schmidt orthogonalization. In the original theory, superscript $^{(b)}$ indicates a frequency band that we denote as domain $(f)$. 
Equation 8 in AHO corresponds to Equation~\ref{hatsf} here.

\begin{figure}
\begin{center}
\includegraphics[height=100mm, width=75mm]{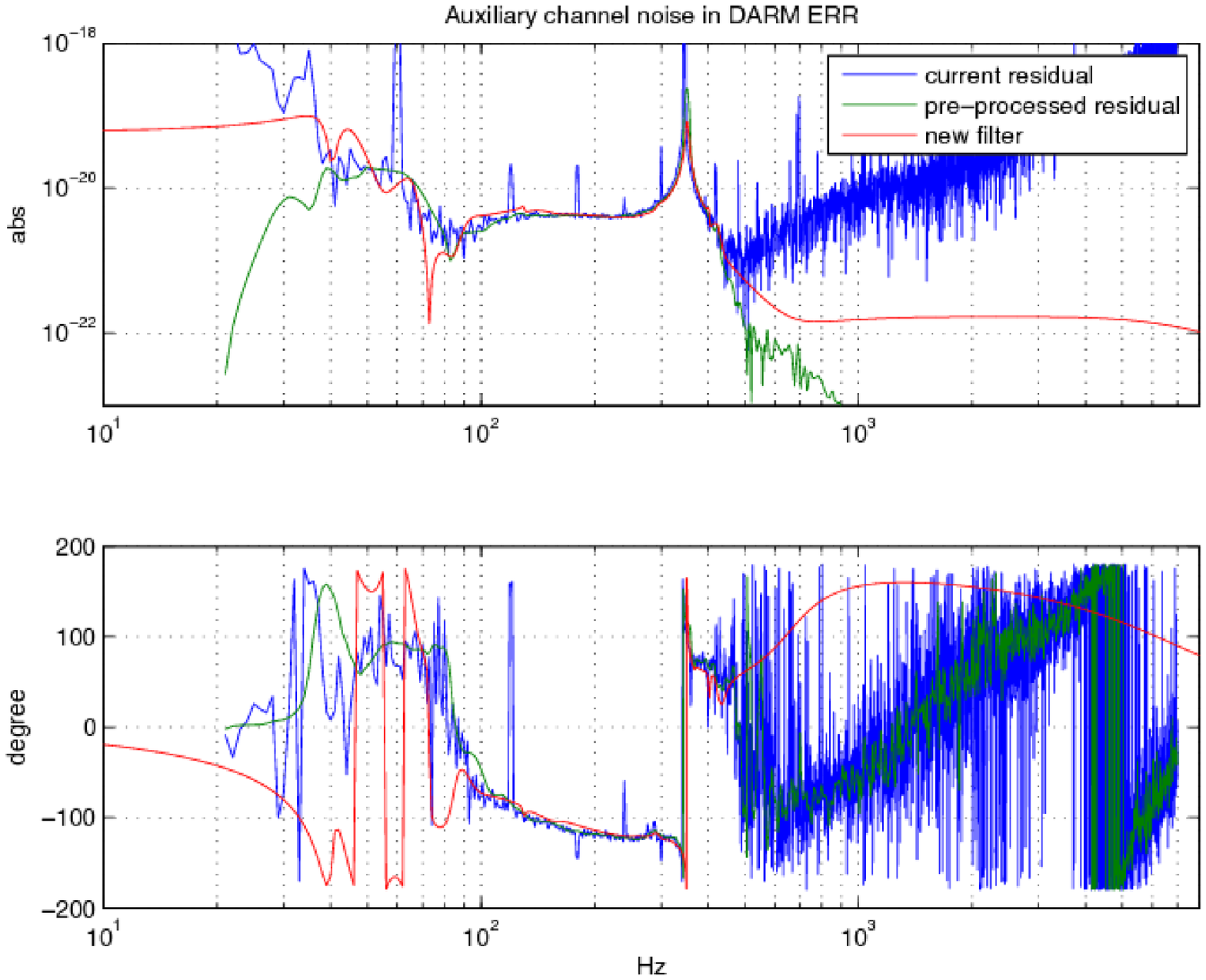}
\includegraphics[height=100mm, width=75mm]{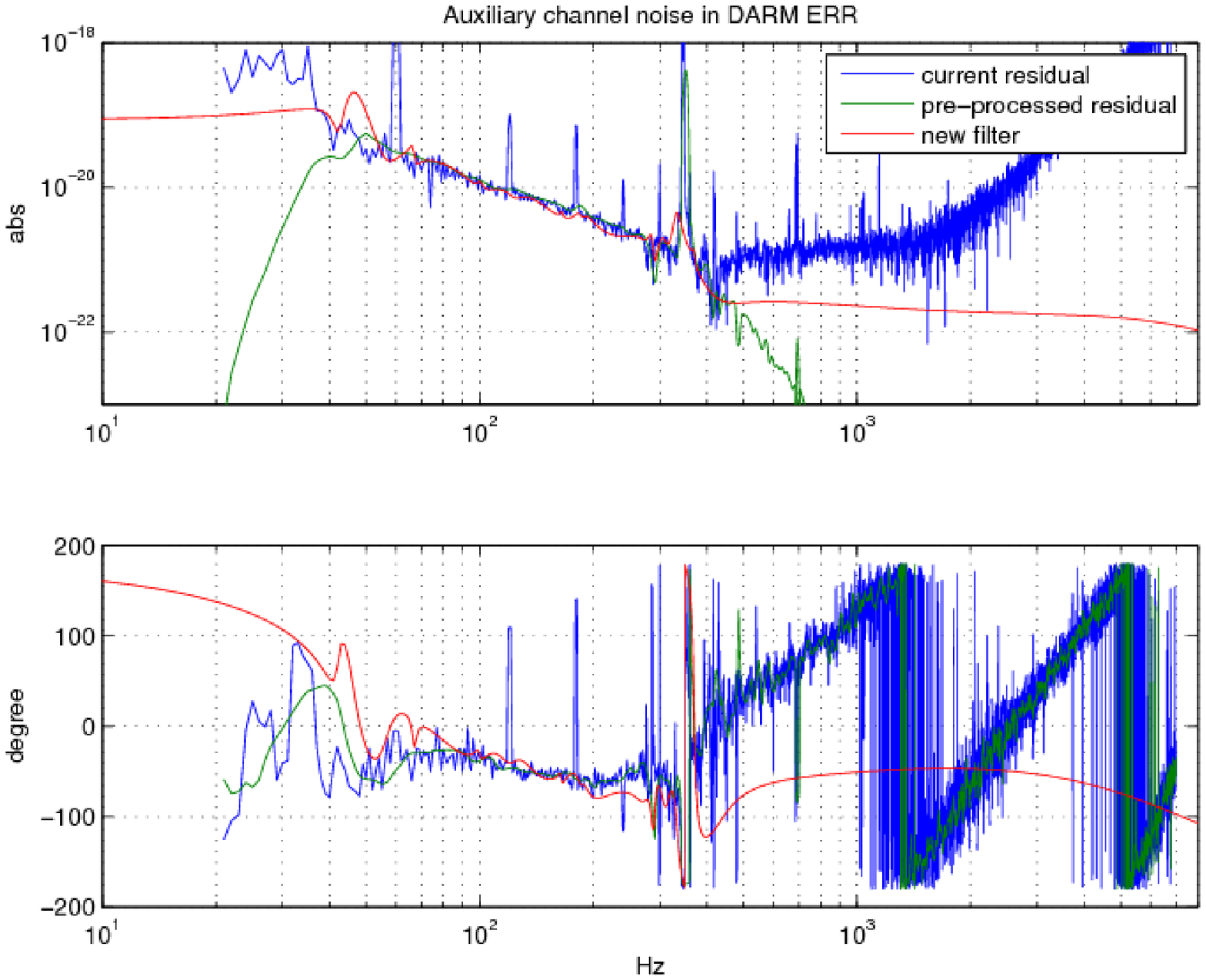}
\caption{Sample transfer function measurements (amplitude and phase) from LIGO Hanford Observatory, H1: 2010 March 21; MICH-$h(t)$ on left, PRC-$h(t)$ on right. Transfer function fit in coherent band -- note the difference between raw data residual and the `pre-processed residual', which has been smoothed and weighted to emphasize known-coherent bands. Units of amplitude spectral density (Hz$^{-1/2}$) and phase (degrees) vs frequency (Hz).}
\label{tfGraph}
\end{center}
\end{figure}

The transfer function $T_{xy}$, from the cross-power ratio of arbitrary channels $x$ and $y$, 
guides the estimated feedforward filter $g$. Figure~\ref{tfGraph} shows the fit to the transfer function. Viewed as an inverse Fourier transform $F^{-1}$, decoupling signal (target, subscript $s$) from noise (witness, subscript $n$):
            \begin{eqnarray}
            g(t) = F^{-1} \left( \textup{fit} \left[ T_{sn} (f) \right] \right) \label{gcf}.
            \end{eqnarray}
\noindent Finally, the post-filtering signal (target) $\hat{s}$ is given by the convolution ($\times$) with $\gamma$, the transfer function coupling noise (witness) into signal (target), $s$ pre-filter signal, $n$ noise, and with channels indexed by $j$ and curly brackets indicating an observable quantity: 
            \begin{eqnarray}
            \hat{s} (t) = \left\{ s + \Sigma_j \left(\gamma_j \times n_j\right)\right\} (t) - \Sigma_j \left(g_{j} (t) \times \left\{ n_{j} \right\} (t)\right) \label{hatsf}.
            \end{eqnarray}

Blind application of this method could produce incorrect noise reduction. 
Application of this paper's method to uncorrelated channels would lead to arbitrary noise reduction by an average analytic factor of $(1 - 1/F)$, where $F$ is the number of bins in a fitted frequency span (equal to the the number of time-domain averages). 
Given 1-s windowing with 50\%-overlap on 1024 s, $F = 2047$, for a false noise reduction of about 0.05\%. 
The ideal of 1024-s windows is not always achievable with LIGO duty cycles. In these cases, AMPS incorporates some filters estimated on as little as 32 s of data, for which the reduction would be 3\%, but only when these filters are averaged together with longer-duration (512 s or greater) filters. No isolated filter uses less than 60 s of data, which could yield a false reduction of 1.5\%. AHO clarify that subtraction is tenable so long as covariance is present at a statistically significant level. They set a benchmark of an order-of-magnitude above the magnitude-square covariance expectation value of $1/F$. Since the AMPS pipeline emphasize fits in regions where the magnitude-squared coherence is greater than 3\%, and often 10\% or more, it usually satisfies their criterion.

\begin{figure}
\begin{center}
\includegraphics[height=100mm, width=75mm]{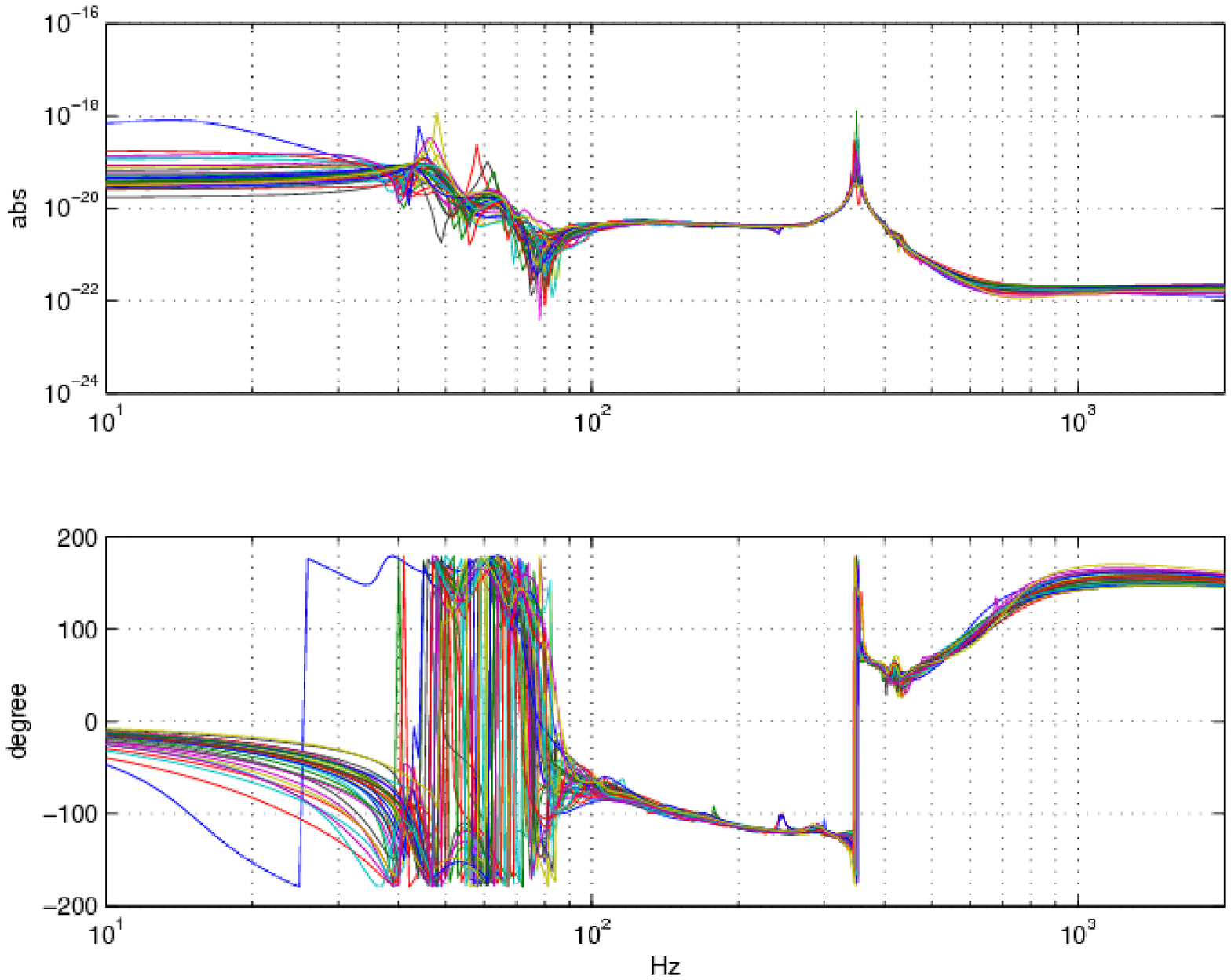}
\includegraphics[height=100mm, width=75mm]{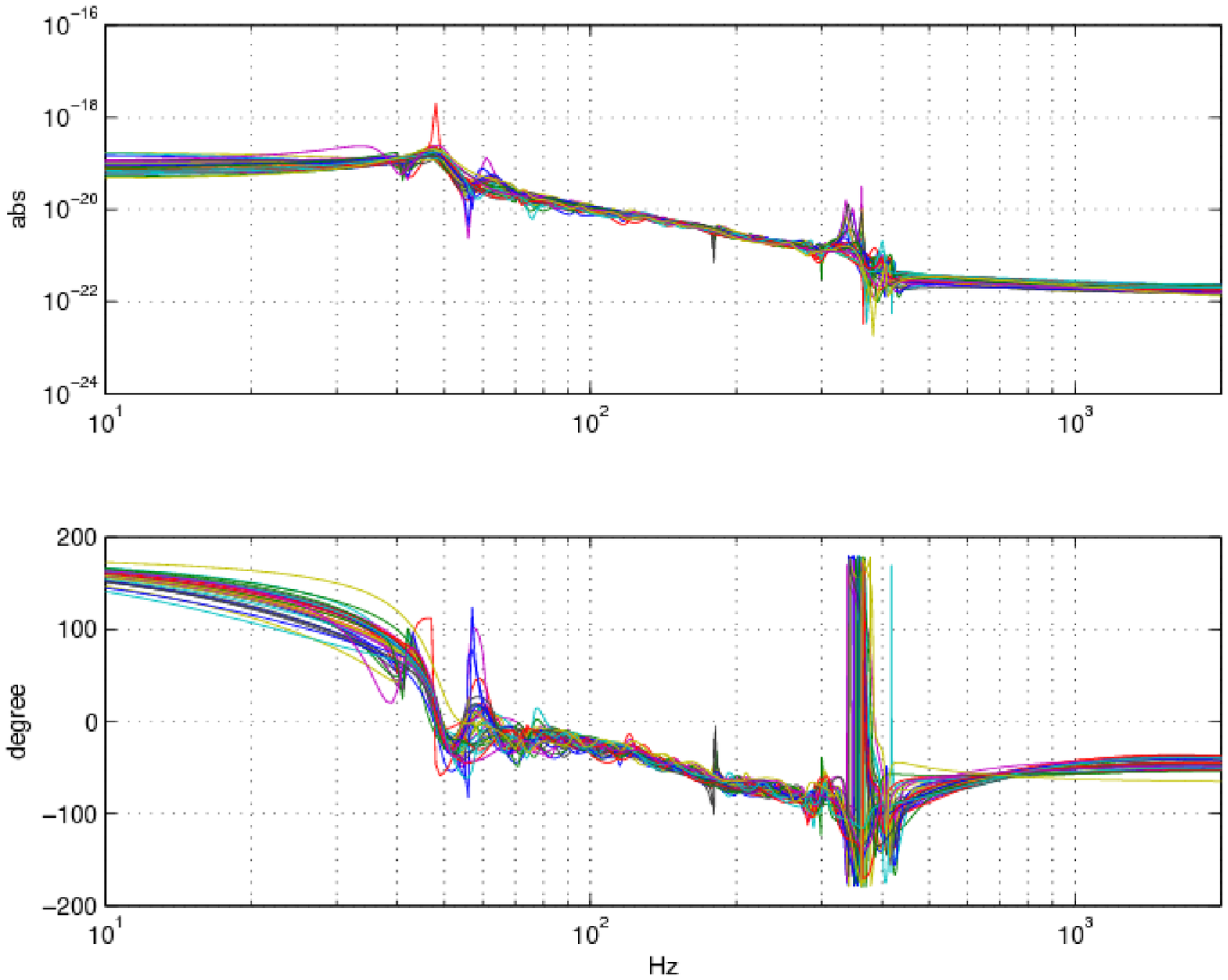}
\caption{Sample Bode plots of fitted ZPK filter functions (amplitude and phase) for multiple 1024 s windows in a science segment, at LIGO Hanford Observatory, H1: 2010 March 21; MICH-$h(t)$ on left, PRC-$h(t)$ on right. Colors only represent different time windows. The similarity in the high-coherence, 80 to 400 Hz band leads us to conclude that the filter design is fairly stable throughout a science segment. Units of amplitude spectral density (Hz$^{-1/2}$) and phase (degrees) vs frequency (Hz).}
\label{BodePlots}
\end{center}
\end{figure}
\begin{figure}
\begin{center}
\includegraphics[height=80mm, width=75mm]{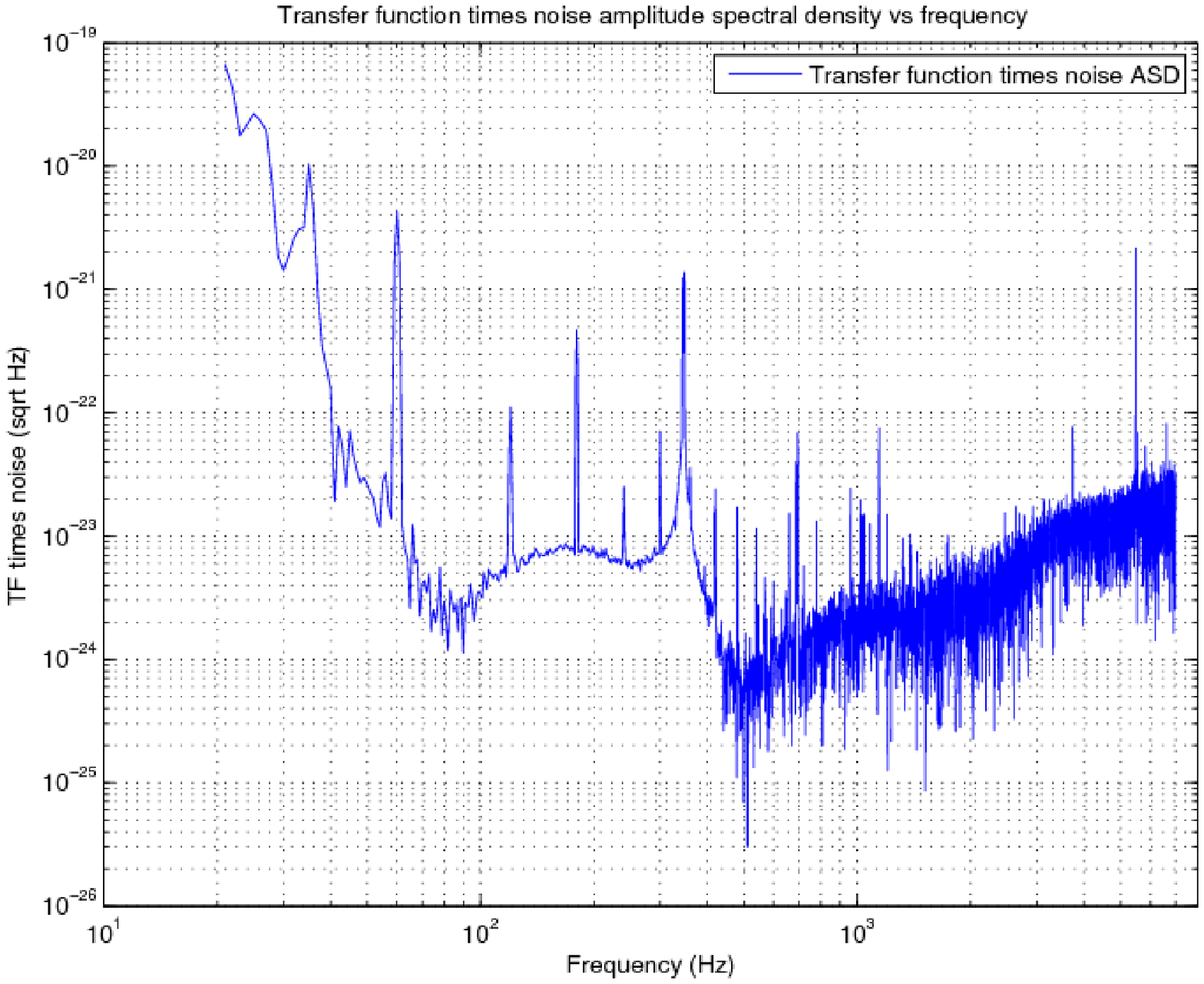}
\includegraphics[height=80mm, width=75mm]{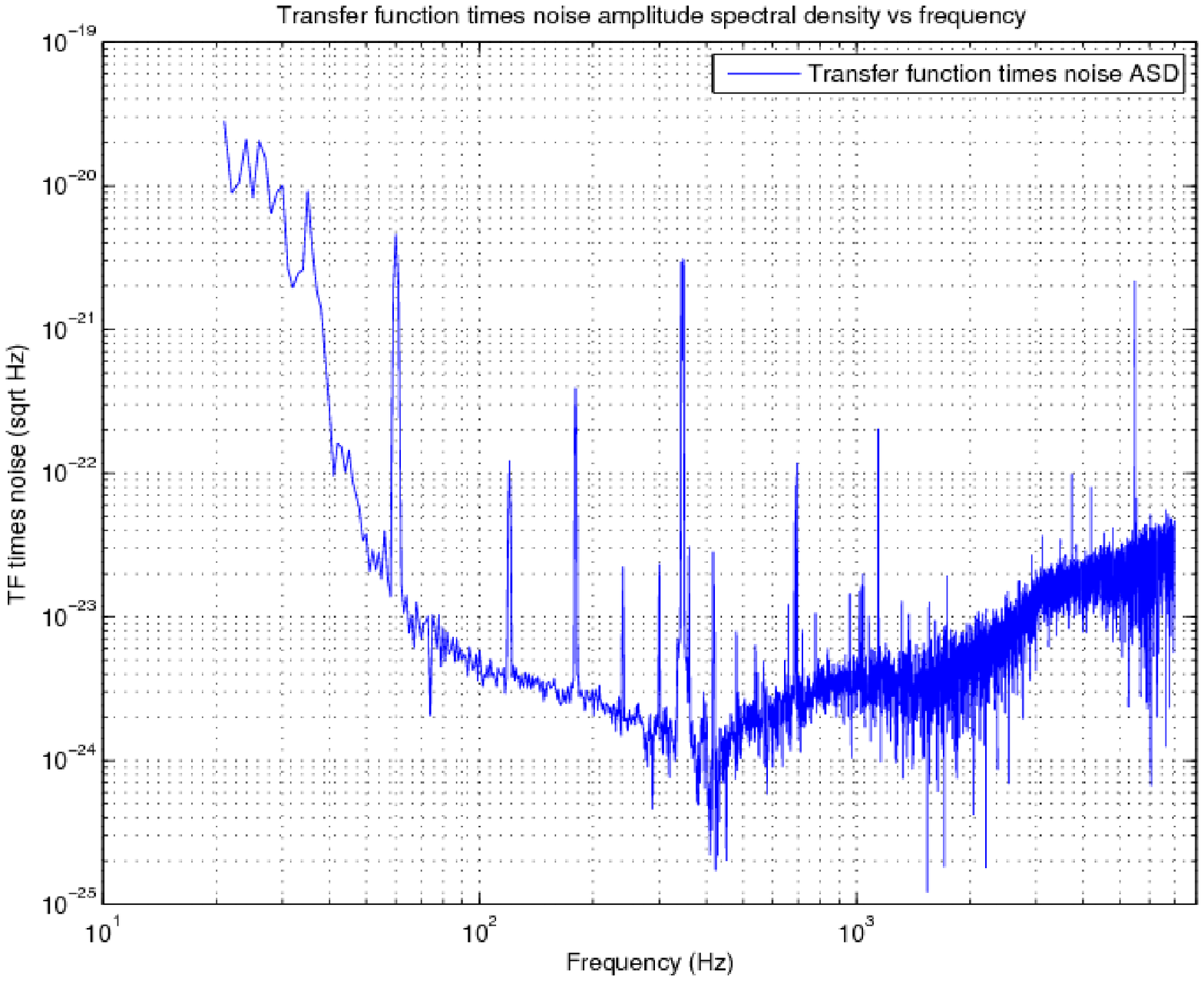}
\caption{Sample subtracted spectra for one window, representing the applied feedforward corrections for each channel during that window, at LIGO Hanford Observatory, H1: 2010 March 21; MICH-$h(t)$ correction on left, PRC-$h(t)$ correction (after MICH-$h(t)$ correction is applied) on right. Units of amplitude spectral density (Hz$^{-1/2}$) vs frequency (Hz).}
\label{subtractedSpectrum}
\end{center}
\end{figure}

As detailed in Section~\ref{filter_fitting-out-of-loop}, filters for each 1024-s window are blended to estimate the target. Figure~\ref{BodePlots} illustrates the short-term consistency of transfer functions during a few-hour science segment, presenting Bode plots of MICH-$h(t)$ and PRC-$h(t)$ transfer function fits for consecutive windows from 2010 March 21 at LIGO Hanford Observatory (magnitude, top and phase, bottom vs frequency [Hz]).  Figure~\ref{subtractedSpectrum} presents sample subtractions. These figures show that AMPS efficiently builds on AHO for operational data for a kilometer-scale gravitational wave interferometer.

Section~\ref{prior_programs} compares transfer function estimators, with Section~\ref{vectfit} being the chosen method. Section~\ref{safeguards} discusses safeguards and vetoes, and Sections~\ref{out-of-loop} and~\ref{filter_fitting-out-of-loop} discuss the details of implementation and verification.

    \section{Feedforward in- and out-of-loop methods}
    \label{prior_programs}
   
        Feedforward subtraction must meet operational constraints. Existing manually-designed filters have long worked, but are more labor-intensive than automated design; new methods, such as Wiener filtering now being considered for seismic and gravity-gradient cancellation~\cite{Driggers2012ActiveNoise} could lead to future improved performance.
        
        \subsection{Manually designed rational filtering in-loop}
        \label{manual_design}

            Manual designs of feedforward functions prove time-consuming. Transfer functions must be manually measured, fit, copied and incorporated into the control system. Additional transfer functions are needed for servo in-loop gain and actuation functions. Manual design is an inefficient choice: it is labor-intensive, and S6 suggests that filter redesign should be performed often.

	While involved, manual-designed rational filtering of MICH and PRC in-loop provides a key part of servo controls to date. Auxiliary controls introduce noise into the DARM channel, so  without real-time correction, the performance would be much worse than design. Most MICH \& PRC subtraction so far comes from real-time corrections; our pipeline makes one to two orders of magnitude smaller corrections.

        \subsection{Frequency-domain automated filter design}
        \label{vectfit}

            AMPS uses Vectfit for periodic re-design. Since the dynamic range in magnitude for transfer functions varies over tens of orders of magnitude, data is pre-processed and weighted to emphasize the most sensitive band.
The method fits an infinite impulse response (IIR) filter onto the witness-to-target transfer function. Since coherence and AHO are linear and transitive, it targets $h(t)$ rather than DARM (noise coupling enters the signal there, but it is wasteful to duplicate the response function).

Each transfer function for a typical 1024 s of data is the average of 1024 independent ratios-of-Fourier-transforms (2047 Hann-windowed, 50\%-overlapping FFTs of 1 s samples). Since FFT error scales with the inverse square root of the number of averages, the relative accuracy is $\mathcal{O}\left(1024^{-1/2}\right)$. The minimum data length, 32 s, yields $\mathcal{O}\left(32^{-1/2}\right)$ relative accuracy. Outside the sensitive band, the fit is deweighted and the transfer function pre-processed, suppressing it by factors of $(f/f_\textup{knee})^\alpha$, where $\alpha = 8$ at low frequencies and $-8$ at high. The $f_\textup{knee}$ values are, respectively, 50 and 400 Hz. AMPS smooths and deweights (Figure~\ref{tfGraph}) known spectral peaks, including 60 Hz harmonics, the LIGO suspension violin modes, and calibration lines. Violin modes are internal resonances of mirror suspensions caused by thermal noise; calibration lines are injections used to track the response function. De-weighting and pre-processsing prevent biasing the filter design with transfer function bands where coherence is low, which would introduce noise. This process leads to convergence with fewer parameters.

 Vectfit converges iteratively, starting with a posited set of poles (32nd order here). About five iterations can converge to a good least-squares-fit for the state-space model, but we require fifteen and complex left-half-plane stability for safety. Root-mean-square (RMS) error, is the threshold for rejecting the filter regression. 
From empirical studies, RMS error above $10^{-18}$ indicates poor fit. 
This test isolates a bad MICH-$h(t)$ correction from a good PRC-$h(t)$ one, or vice versa. 
	We fit one channel at a time, as Section~\ref{out-of-loop} discusses. After this test, the transfer function model is extracted. 

Zero-pole-gain (ZPK) format is used to trim out-of-band zeroes and poles and multiply by a 2nd order Butterworth low-pass filter just below the Nyquist frequency of 8192 Hz, placing poles at 7 kHz to keep causality. A scale factor keeps the filter gain at 150 Hz the same value as without the low-pass filter. Then the ZPK model is refactored into second-order-sections (SOS) for numerical stability. Instead of the inverse Fourier transform of Equation~\ref{gcf}, the model is converted from continuous time (or $s$-domain) to discrete time (or $z$-domain). 

Each filter is applied to its respective witness: the estimated true $h(t)$ target equals the original $h(t)$ measurement minus the filtered witnesses. 
This procedure assumes that coupling from each witness into $h(t)$ is linear. Further, it assumes that 2nd-order coupling, from one witness into the other, is negligible (we estimate the relative contributions to be $\mathcal{O}(10^{-5})$). Spot-checks confirm that these simplifications are justified.  


        \subsection{Wiener filters}
        \label{wiener_filters}

            Wiener filtering~\cite{Wiener1949} would give an optimal filter that minimizes the squared error of the residual, for all the spectrum. Low-frequency MS-coherence of MICH \& PRC with $h(t)$ is small, but Wiener fits them due to high RMS error in that band. Filtering at high RMS power, such as the seismic and Newtonian gravity gradient bands, can allow Wiener filtering directly, but MICH \& PRC would need other methods. Noise whitening~\cite{Driggers2012NN,DeRosa2012FF} uses cost functions to limit out-of-band noise. Wiener filtering sub-spectra could also circumvent contamination, as with wavelet transforms~\cite{KlimenkoSite}.  


        \subsection{Prospects for near-real-time filtering}
        \label{real-time}
      
            AMPS runs a few times faster than real-time on a single 2013 CPU core whilst conducting tests and safeguards, documented in Section~\ref{safeguards}. 
The minimal time lag for a modal sample is one window (1024 s), undesirable for electromagnetic follow-up and multi-messenger astronomy~\cite{Swift2012,Antares2013} sought for Advanced LIGO. 
Such speed is acceptable for secondary $h(t)$-reconstruction when another $h(t)$ exists but inadequate for in-loop, real-time production. 

Near-real-time filtering might be useful for countering upconversion and non-linear cross-coupling, using recent data for training sets, but this is not yet implemented. Until then, the existing method can generate a filter as-needed for real-time use, as prototyped on H1 in the last month of S6.

\section{Safeguard and veto methods}
\label{safeguards}


  \subsection{Calibration integrity}
  \label{CalibrationSection}

It is of vital importance for our noise subtraction scheme to keep the integrity of the calibration of LIGO instruments intact.
If the witness channel contains a cross coupled term proportional to the differential arm length motion $\delta(L_-)$, this term is subtracted from the target and thus could change the calibration of the target in theory.
A simple calculation shows that, for a known coupling mechanism, this effect is on the order of $10^{-5}$ for LIGO and other similarly configured instruments.

Even without considering any feedback control mechanism, theoretically MICH and DARM signal both have cross contamination terms proportional to $\pi/2\mathcal{F}$~\cite{SiggFreq1997}:
\begin{eqnarray}
\textup{DARM} \propto \delta(L_-) + \frac{\pi}{2 \mathcal{F}} \delta(l_-) \label{CalDARMdef}, \\
\textup{MICH} \propto \delta(l_{-,0}) + \frac{\pi}{2 \mathcal{F}} \delta(L_-) + n \label{CalMICHdef},
\end{eqnarray}
where $\mathcal{F}$ is the finesse of the arm cavities, $n$ the sensing noise of the MICH, and $\delta(l_{-,0})$ the natural fluctuation of the Michelson path difference caused by seismic motion etc.
Since  $\delta(l_{-,0})$  is not coherent with $ \delta(L_-)$, and since MICH is dominated by $n$ in our frequency band of interest, we can ignore $\delta(l_{-,0})$ in this discussion.

The servo system with an open loop transfer function of $G_M$ to keep MICH from going out of linear range would inject $\frac{\pi}{2 \mathcal{F}} \delta(L_-) + n$ term to the physical Michelson length difference $\delta(l_-)$, which in turn affects the DARM signal:
\begin{eqnarray}
\delta(l_{-}) = - \frac{G_M}{1+G_M} \left( \frac{\pi}{2 \mathcal{F}}\delta(L_-) + n \right) \label{offsetMICHsignal}\\
\textup{DARM} \propto \delta(L_-) -\frac{\pi}{2 \mathcal{F}} \frac{G_M}{1+G_M} \left( \frac{\pi}{2 \mathcal{F}}\delta(L_-) + n \right). \label{offsetDARMsignal}
\end{eqnarray}
The DARM signal, uncorrected by feedforward, is now contaminated by MICH noise term $n$ as well as a small correction term for $ \delta(L_-)$, both due to the MICH feedback. Note that $G_M/(1+G_M)$ is on the order of 1 or smaller, and  $(\pi/2 \mathcal{F})^2$ is on the order of $10^{-5}$, so the correction term is on the order of $10^{-5}$ or smaller.

Feedforward subtraction looks at the MICH control signal,
\begin{equation}
\textup{MICH}_{ctrl} \propto \frac{\pi}{2\mathcal{F}}\delta(L_-) +n,\label{offsetMICHCTRL}
\end{equation}
and subtracts $n$ from DARM, and in the process also subtracts $\frac{\pi}{2\mathcal{F}}\delta(L_-)$. Looking at Eqs.~\ref{offsetDARMsignal} and~\ref{offsetMICHCTRL}, as far as the noise reduction is observed in DARM, the small $\delta(L_-)$ term is also reduced, and the impact on the calibration is on the order of $(\pi/2 \mathcal{F})^2=O(10^{-5})$ at most.

A similar argument can be made for PRC, but this time the cross coupling is not only dependent on $(\pi/2\mathcal{F})$ but also on the asymmetry of the arms, further reducing the coupling.

Note that the above mentioned discussion is equally applicable to real time as well as post-facto feedforward. LIGO uses real time feedforward, and the online feedforward described in this paper subtracts only a small amount of noise left uncaught by the real time system. VIRGO implements similar real time feedforward to remove 50Hz line successfully~\cite{Buskulic2000}. In both of these cases, no measurable effect caused by feedforward has been reported.

Nevertheless, two checks were performed to see if there is any unknown mechanism to compromise the calibration of DARM, which are explained in Section~\ref{postprocessingSafeguards}.

\subsection{Runtime safeguards}
    \label{runtimeSafeguards}

Safeguards and vetoes then verify data integrity against possible issues. These issues include degrading data, offsetting and incorrectly time-stamping the data, incorrectly subtracting $h(t)$ from itself, and introducing windowing artifacts.

Amplitude spectral density (using Welch's method~\cite{Welch1967}) leads to an estimate of inspiral range $\mathcal{R}$. Inspiral range~\cite{FinnInspiral1993} in LIGO detector characterization refers to the orientation-and-direction-averaged distance at which a 1.4-1.4 solar mass neutron star binary coalescence could be detected with an SNR of 8. A window's filtered data is used only if it passes two cuts. The post-filter $\mathcal{R}$ must be at least 99.9\% of unfiltered $\mathcal{R}$. None of the 40 points in a `comb' (each point being 5/16 Hz wide) of quiet bands can be noisier than 1.2 times uncorrected $h(t)$. The factors are chosen empirically to permit expected noise fluctuations; most surviving data is superior.  

If cuts are triggered, the filter is rejected. To avoid discontinuities and add robustness against non-stationarity, the windowing procedure (Section~\ref{filter_fitting-out-of-loop}) is re-run to merge successfully-filtered data smoothly with unfiltered $h_0$. Cut tests are also re-run; if passed, the data is used, else the unfiltered data is progressively weighted further for eight more attempts. If all fail, the final attempt is written and the program proceeds. Empirically, almost all written data is an improvement.

    \subsection{Post-processing safeguards}
    \label{postprocessingSafeguards}

\begin{figure}
\begin{center}
\includegraphics[height=75mm, width=150mm]{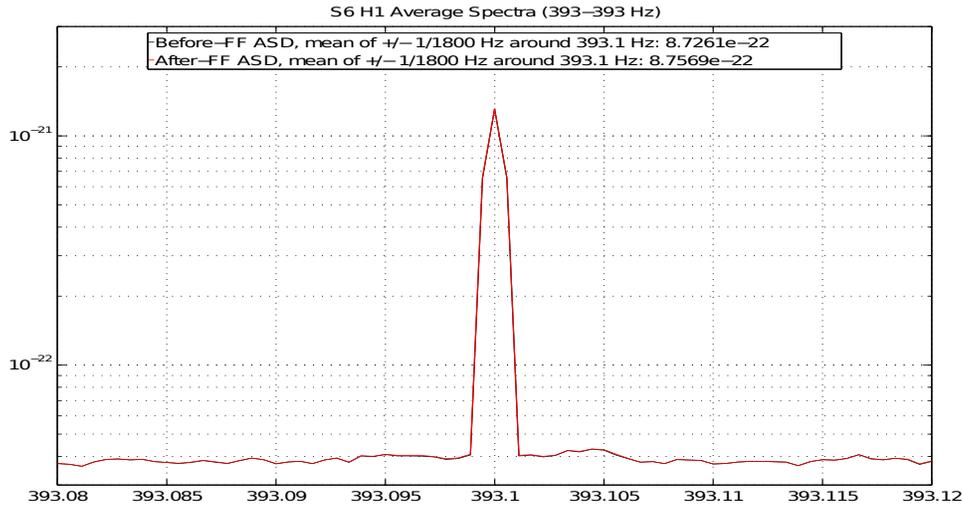}
\caption{Calibration line test: before-feedforward mean of the 393.1 Hz line and two neighboring FFT bins was $8.7261 \times 10^{-22}$, after was $8.7569 \times 10^{-22}$. Feedforward made the calibration line region noisier by $3.1 \times 10^{-24}$ or 0.35\%, suggesting that we correctly apply Hann-windowed feedforward without subtracting true $h(t)$. Moreover, no spectral line combs are observed to either side of the calibration line peak at 393.1 Hz, indicating that the method does not introduce windowing artifacts.}
\label{calLineTest}
\end{center}
\end{figure}

\begin{figure}
\begin{center}
\includegraphics[height=75mm, width=150mm]{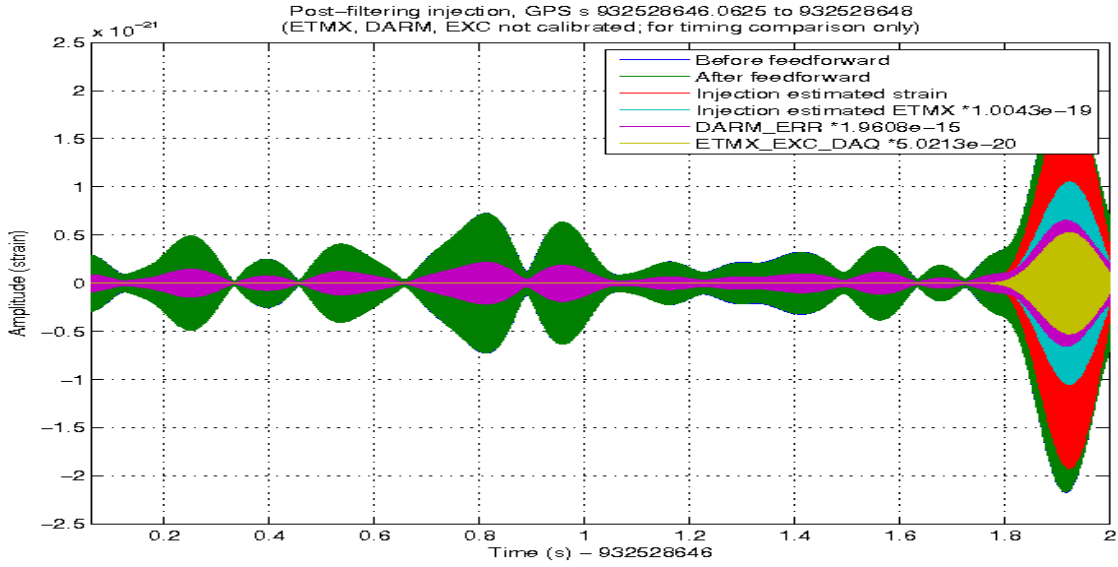}
\caption{Time-domain plot of diagnostic channels from a burst injection. Colors are illustrative only to the fact that the envelopes of the traces increase after 1.8 s, indicating that the burst injection time is correct in the new data. `Before feedforward' and `after feedforward' traces occult each other in the graph, because they are almost identical. `Before feedforward' is $h(t)$ data; `after feedforward` is $h(t)$ with feedforward subtraction. `Injection estimated strain' is the digital injection as intended to be introduced into strain, but the actual injection is made on the end test mass X (ETMX), so the calibrated `Injection estimated ETMX' is also displayed. Raw `DARM\_ERR' and `ETMX\_EXC\_DAQ' are redundant but reinforce the trend.}
\label{timeDomainInjection}
\end{center}
\end{figure}

\begin{figure}
\begin{center}
\includegraphics[height=75mm, width=150mm]{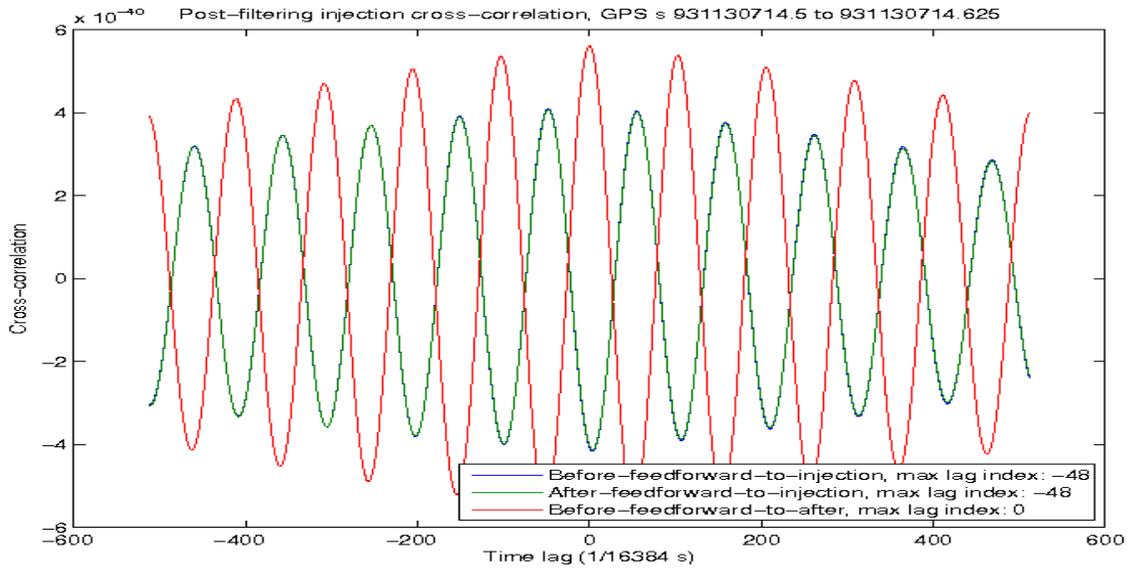}
\caption{Cross-correlation pairwise between $h(t)$ pre-, post-feedforward, and ETMX injection data: the extrema and zero-crossings match.  Note both before-feedforward (blue) and after-feedforward (green) strain traces are almost identical and therefore overlap. The strains appear inverted, but in the same way, due to a sign error in the hardware injections at this time. The absence of a time lag shift between before and after indicates that feedforward has not altered the phase of the data, at least for this injection. The equivalence in cross-correlation magnitude indicates that amplitude also is unaffected.}
\label{crossCorrelationGraph}
\end{center}
\end{figure}

Diagnostics check whether 
calibration lines (Figure~\ref{calLineTest}) are preserved
and 
injections (Figures~\ref{timeDomainInjection} and \ref{crossCorrelationGraph}) are recovered.

Calibration line studies seek to answer two questions: whether feedforward distorts the signal or adds noise. 

In post-processing, `Short Fourier Transforms' (SFTs) were made with a frequency resolution (1/1800 Hz) from corrected $h(t)$. These SFTs are much shorter than the science run; 1800 s is standard for continuous wave searches. 
Signal distortion is evaluated using the mean of three bins in [393.1 - 1/1800, 393.1 + 1/1800] Hz.
Evaluating multiple bins accounts for some spectral leakage; the bin-centered central line is much larger. 
For the $10^6$ s of H1 science time analyzed, the before-feedforward mean was $8.7261 \times 10^{-22} (\textup{Hz})^{-1/2}$, whereas after it was $8.7569 \times 10^{-22} (\textup{Hz})^{-1/2}$. Feedforward made the calibration line region noisier by $3.1 \times 10^{-24} (\textup{Hz})^{-1/2}$ or 0.35\%, perhaps due to a deweighted fit around the 393.1 Hz line, to avoid biasing more sensitive parts of the spectrum. It does not affect the calibration, since it is consistent with MICH and PRC leakage merely raising noise floor near the line. We also check the calibration line at 1144.3 Hz (before: $3.1190 \times 10^{-20} (\textup{Hz})^{-1/2}$, after: $3.1188 \times 10^{-20} (\textup{Hz})^{-1/2}$), which is actually less noisy. The 46.7 Hz line is too low-frequency to measure with these SFTs. 

To test for noise addition, we searched for windowing artifacts, e.g., spectral combs with spacing of 1/1024 or 1/512 Hz, around a prominent line.
No new combs or other artifacts were obvious in our 1800-s, 50\%-overlapping Hann-windowed SFTs before/after comparison of approximately $10^6$ s of H1 science time between GPS times $931.0 \times 10^6$ (2009 July 07) and $932.8 \times 10^6$ (2009 July 28), 
focused on the 393.1 Hz calibration line. Strictly speaking, the line visible in $h(t)$ is a residual from imperfect cancellation of control and error signals used in $h(t)$ construction from DARM error and control signals. The nature of the line does not affect our analysis, because neither MICH nor PRC contain or affect it.

Injection studies first examined compact binary coalescence and sine-Gaussian injections~\cite{LIGOBurst2012} for GPS seconds $931.0 \times 10^6$ (2009 July 07) to $932.8 \times 10^6$ (2009 July 28). We then calculated the matched-filter SNR of each injection in S6. Each SNR is directly proportional to the distance at which such a signal \textit{can} be observed, and therefore also to the instrumental sensitivity to signals of that type~\cite{Findchirp2012,PetersMatthews1963}. Higher SNR (mean 3.99\% H1, 2.77\% L1) was found. Effective distance as \textit{recovered} is inversely proportional to signal, appeared nearly unchanged in these injections (mean -0.00347\% H1, +0.307\% L1), establishing that the SNR increase came from N decreasing rather than S increasing. The constant injection effective distance reinforces that the calibration is unchanged. These tests affirmed that feedforward data contains recoverable (slightly higher SNR ratio) injections at the correct time and phase. 

We conclude that we are not subtracting $h(t)$ from itself.

    \section{Feedforward with MICH and PRC channels}
    \label{out-of-loop}

\begin{figure}
\begin{center}
\includegraphics[height=80mm,width=150mm]{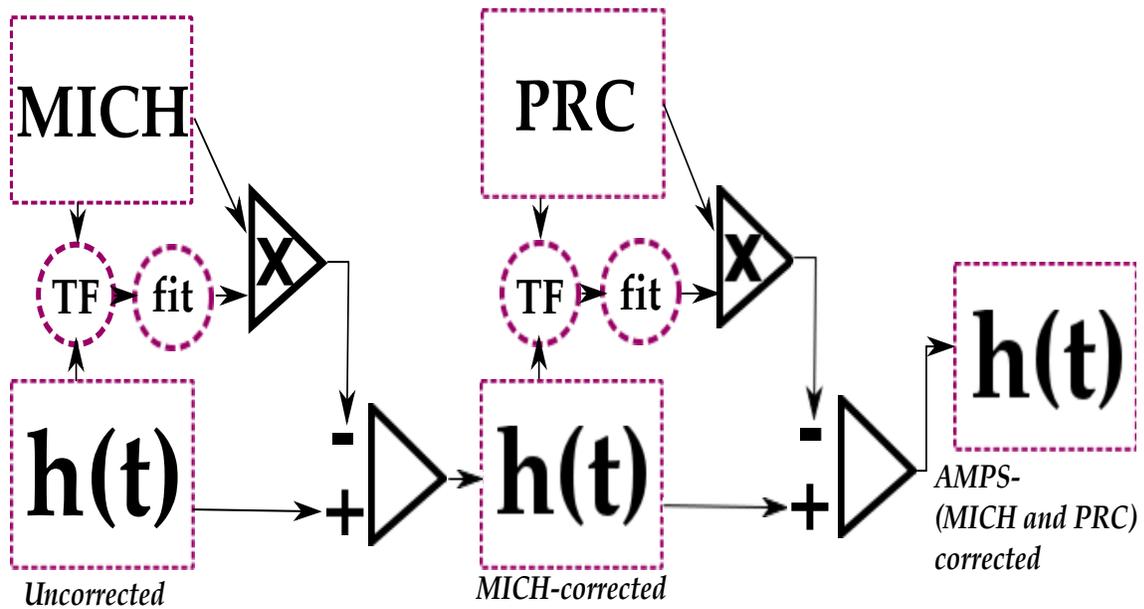}
\caption{Feedforward subtraction pipeline to read in $h(t)$, MICH, PRC, and write out AMPS-corrected $h(t)$. Data flows schematically from left to right; the MICH-$h(t)$ stage output is used as input for the PRC-$h(t)$ stage, then data is written. Code online: \texttt{http://ligo-vcs.phys.uwm.edu/wsvn/MatApps/packages/detchar/AMPS/}\newline \texttt{trunk/aletheia.m}}
\label{pipelineGraph}
\end{center}
\end{figure}    

A post-facto, linear filter is fitted and applied to either MICH or PRC (serially). Fits occur in the frequency-domain, and application occurs in time-domain. Reading in $h(t)$, correcting it with MICH and PRC, and writing the result (including data quality and state vectors), the pipeline is shown in Figure~\ref{pipelineGraph}.

        \subsection{Filter fitting across science segments}
        \label{filter_fitting-out-of-loop}

\begin{figure}
\begin{center}
\includegraphics[height=110mm,width=150mm]{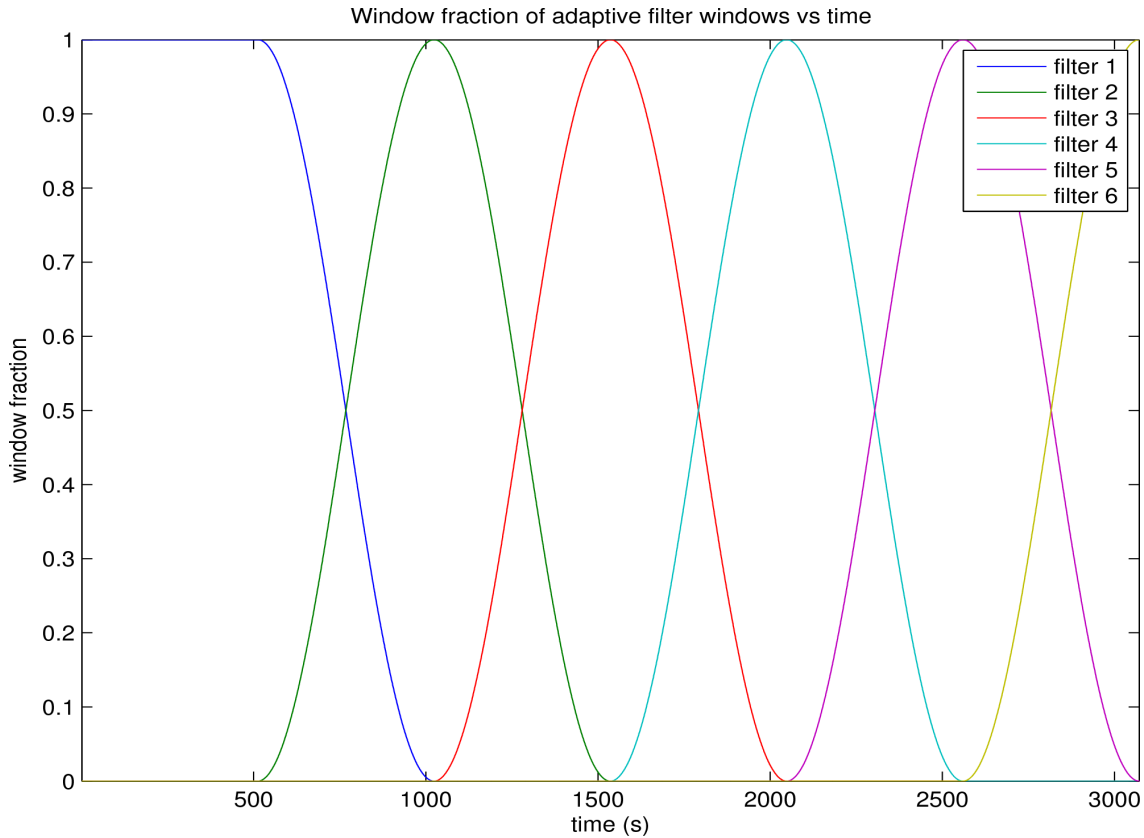}
\caption{Schematic windowing for one LIGO science segment, illustrating windowing after an initial half-window offset. Filters are calculated for windows up to 1024-s, then 50\%-overlapping Hann windows merged, giving a corrected measurement of $h(t)$. Code online: \texttt{http://ligo-vcs.phys.uwm.edu/wsvn/MatApps/packages/detchar/AMPS/}\newline \texttt{trunk/eleutheria.m}}
\label{windowingScience}
\end{center}
\end{figure}

We run one job process per science segment. Segments range in duration from seconds to days, with a median of hours. The interferometer is locked during each segment, meaning it is held fixed on a fringe, by servo control. Lock loss or noise degradations can define a segment end. Segments are divided into 50\%-overlapping Hann windows, filtered, and smoothly re-merged. Windowing is idempotent for $h(t)$ itself; the difference from window to window is the correction added. The first 512 s derive only from the first window; every 512 s afterward, a new window commences, as in Figure~\ref{windowingScience}.

Using Equation~\ref{hatsf} with filters $g$, target $S = \left\{ s + \Sigma_j \left(\gamma_j \times n_j\right)\right\}$ and witness $N_j = \left\{ n_{j} \right\}$, we can evaluate  $\hat{s} (t)$. Since the filters for different channels are calculated in series, with transfer functions $T$, Equation~\ref{hatexpand} has $g_1 \sim T_{S,N_1}$ but $g_2 \sim T_{\left( S-g_1 \times N_1 \right),N_2}$. Here, $S$, $N_1$ and $N_2$ are respectively $h(t)$, MICH and PRC.

            \begin{eqnarray}
            \hat{s} (t) &=& S (t) - \Sigma_j \left(g_{j} (t) \times N (t)\right) \label{hatexpand}, \\
		&=& S (t) - g_1 (t) \times N_1 (t) - g_2 (t) \times N_2 (t) \label{hatexpanded},\\
&\sim& S (t) - F^{-1} \textup{fit}\left[ T_{S,N_1} \right] \times N_1 (t) - F^{-1} \textup{fit}\left[T_{\left( S-g_1 \times N_1 \right),N_2} \right] \times N_2 (t).
            \end{eqnarray}

	Since $N_1(t)$ and $N_2(t)$ are added linearly to $S(t)$, we can analyze them independently. 
Analyze the first two terms of Equation~\ref{hatexpanded} and take $N(t) = N_1(t)$. Let $g_{A}$ and $g_{B}$ be the earlier and later filters for $N(t)$ being time-domain merged in a Hann-window; they are respectively calculated from overlapping data sets $[S_A, N_A]$ and $[S_B, N_B]$. The sets are identical at time $t$, so $S(t)=S_A(t) = S_B(t)$, $N(t)=N_A(t) = N_B(t)$. Windowing merges data streams $\hat{s}_A$ and $\hat{s}_B$ over $\tau = 1024$ s, per Equation~\ref{windowexpand}: 

	\begin{eqnarray}
        \hat{s} (t) &=& \frac{\hat{s}_A (t)}{2} \left[1 - \cos \frac{2 \pi (t+\frac{\tau}{2})}{\tau} \right] + \frac{\hat{s}_B (t)}{2}\left[1 - \cos \frac{2 \pi (t+\tau)}{\tau} \right] \label{windowexpand}, \\
	  &=& \frac{1}{2} \left( \hat{s}_A (t) + \hat{s}_B (t) + \cos \frac{2 \pi t}{\tau} \left[\hat{s}_A (t) - \hat{s}_B (t) \right]\right),\\
	&=& \frac{2 S(t) - (g_{A} + g_{B})\times N(t)}{2} - \frac{g_{A} - g_{B}}{2} \times N(t) \cos \frac{2 \pi t}{\tau},\\
          &=& S(t) - \frac{1}{2} \left( g_{A} \left [1 + \cos \frac{2 \pi t}{\tau} \right] + g_{B} \left [1 - \cos \frac{2 \pi t}{\tau} \right] \right) \times N (t) \label{windowexpanded}.
	\end{eqnarray}

	Equation~\ref{windowexpanded} shows that the windowing process equates to evolving filter coefficients with a 512 s cadence. Substitute $\hat{s}(t)$ into $S(t)$ with $N(t) = N_2(t)$ for the next noise channel to extend the result.

            Science segments are subdivided into at most 16384 s. These subdivisions overlap for 512 s, so each side calculates identical filters for the overlap, but only the latter half writes the overlap, to avoid race conditions. Label $g_W, g_X, g_Y, g_Z$ the final filters calculated in job 1; $g_A, g_B, g_C, g_D$ are the first in job 2. Where each parenthesis contains 512 s and the addition sign denotes Hann-windowing of the filters, the end of job 1 is $\ldots(g_W+g_X)(g_X+g_Y)(g_Y+g_Z)$ and the start of job 2 is $(g_A+g_B)(g_B+g_C)(g_C+g_D)$. Overlap denotes that filter $g_A$ is derived from the same data as filter $g_Y$, and likewise $g_B \simeq g_Z$. Thus $(g_A+g_B)=(g_Y+g_Z)$. Segments shorter than 60 s are not filtered, and dangling windows shorter than 32 s are rolled into their predecessors. These provisions prevent filters based on insufficient data. 

    \section{Results of feedforward}
    \label{all-results}

        \subsection{Post-processing diagnostics}
        \label{diagnostics}

\begin{figure}
\begin{center}
\includegraphics[height=120mm, width=150mm]{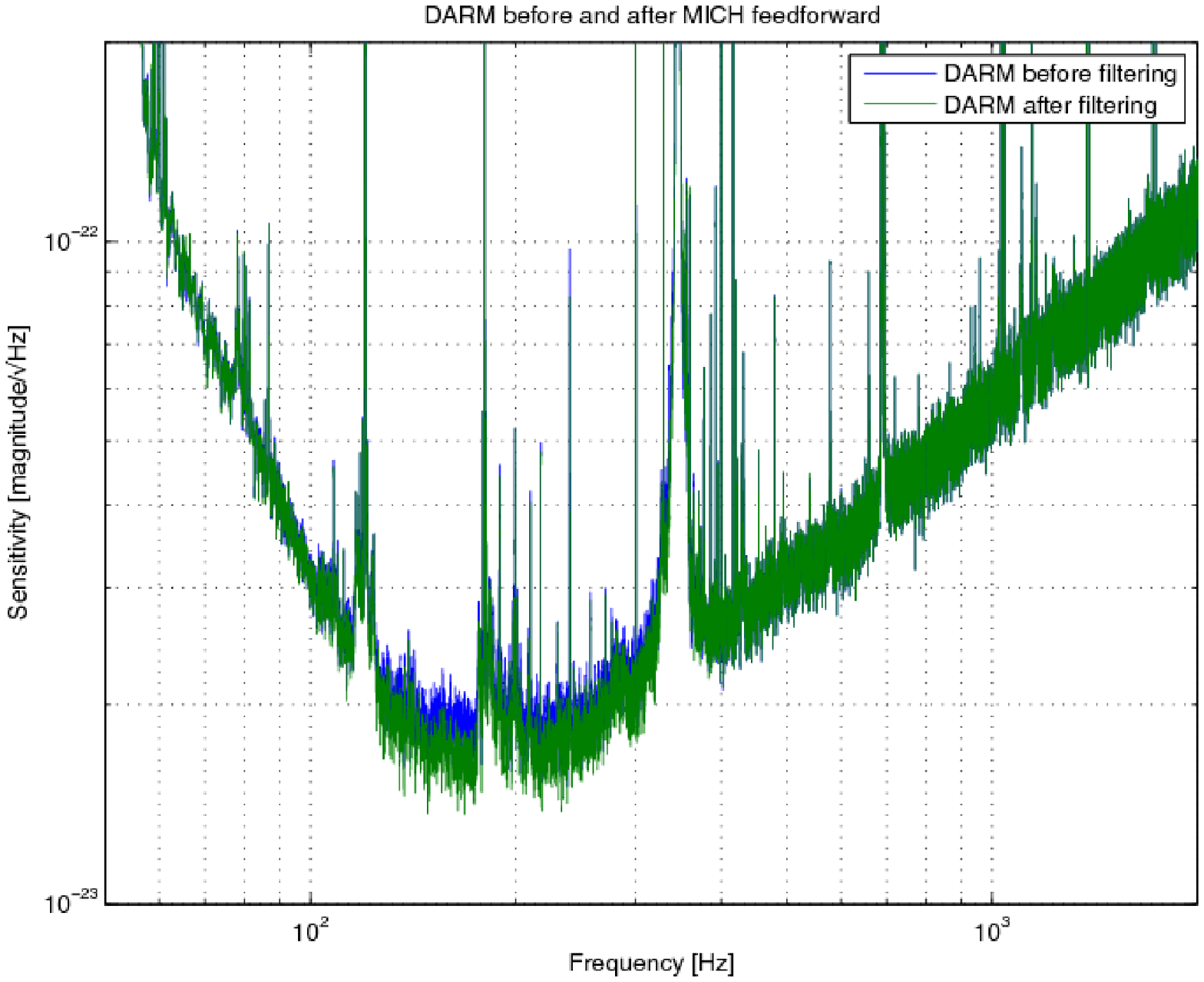}
\caption{Exemplar of a typical case, +1.1 Mpc (5.9\% inspiral range)
\textit{(GPS time 953164819 to 953165839, 2010 March 21)}. Read `DARM' as $h(t)$, `MICH' as `MICH-PRC'. The most benefit is seen in the 80 to 400 Hz band, especially around 150 Hz, where LIGO is most sensitive. The main fundamental limit in this band is thermal suspension noise, but historically auxiliary channel noise has been a major contaminant. Note that the 60 Hz and harmonic lines are not subtracted, although a separate magnetometer servo does reduce their impact. The 330 Hz violin mode is not likely amenable to feedforward.}
\label{typicalInspiralGraph}
\end{center}
\end{figure}

\begin{figure}
\begin{center}
\includegraphics[height=120mm, width=150mm]{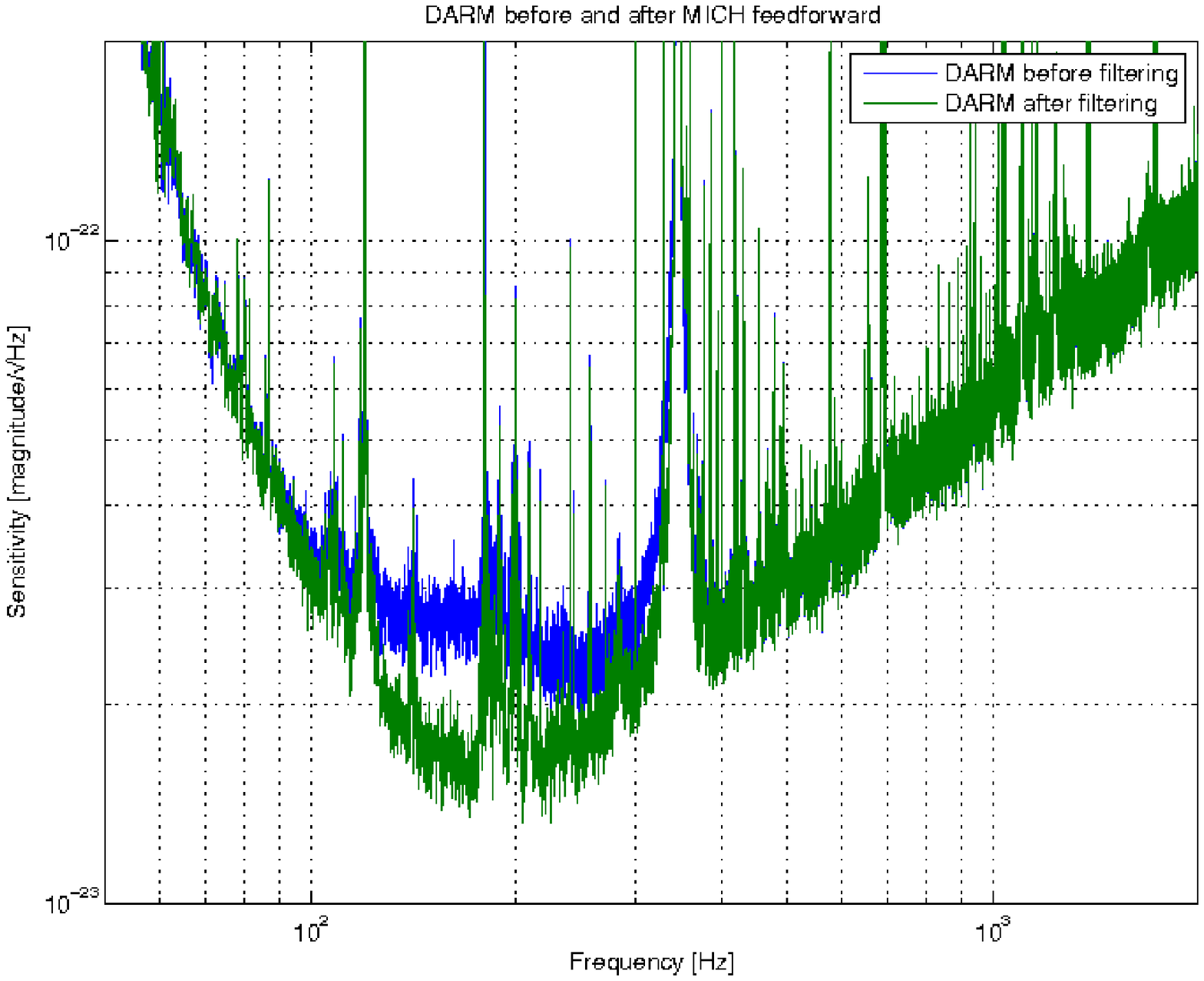}
\caption{Best improvement seen in S6 for H1 $h(t)$, +4.4 Mpc (29\% inspiral range)
\textit{(GPS 955187679 to 955188191, 2010 April 13)}. Read `DARM' as $h(t)$, `MICH' as `MICH-PRC'. Such a loud cross-coupling would be noticed in real-time by the on-site staff. The elevated noise floor is unusual in science mode, but the fact that feedforward corrects it suggests the importance of controlling auxiliary channels to prevent such glitches. The post-feedforward spectrum is comparatively normal for science mode, as in Figure~\ref{typicalInspiralGraph}.}
\label{bestInspiralGraph}
\end{center}
\end{figure}

Lower spectral noise floors reveal improvement in Figures~\ref{typicalInspiralGraph} and~\ref{bestInspiralGraph}. Feedforward most improves the spectra with elevated noise levels. It enhances sensitivity less when the interferometer is already optimized. This tendency is consistent with underlying thermal and shot noise. Glitches contaminate $h(t)$ less when the servo-to-strain coupling is minimized. Insofar as coupling non-stationarity evolves slower than the 512-s windowing timescale, adaptive filtering appears to reduce the impact of glitches.


\begin{figure}
\begin{center}
\includegraphics[height=100mm, width=75mm]{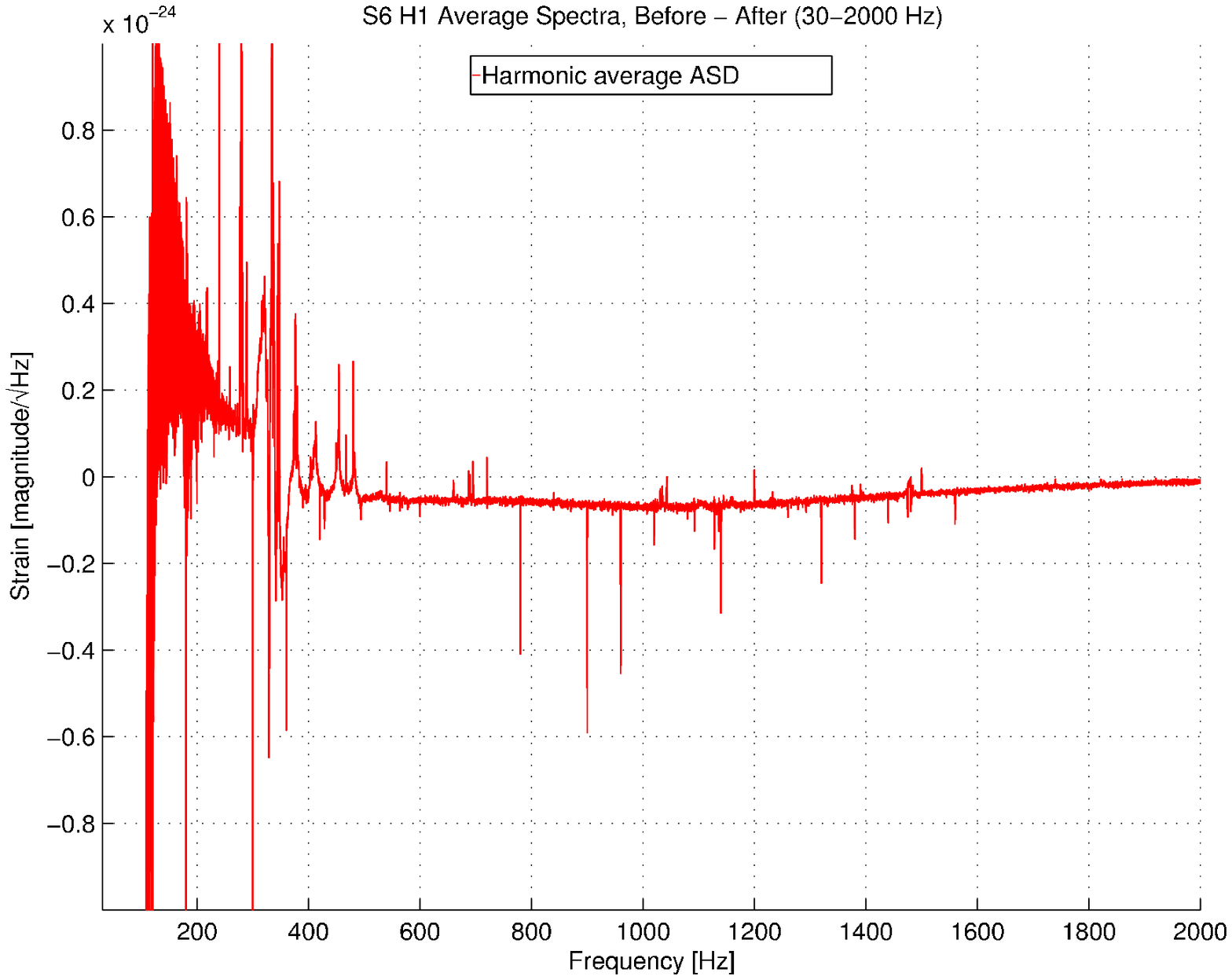}
\includegraphics[height=100mm, width=75mm]{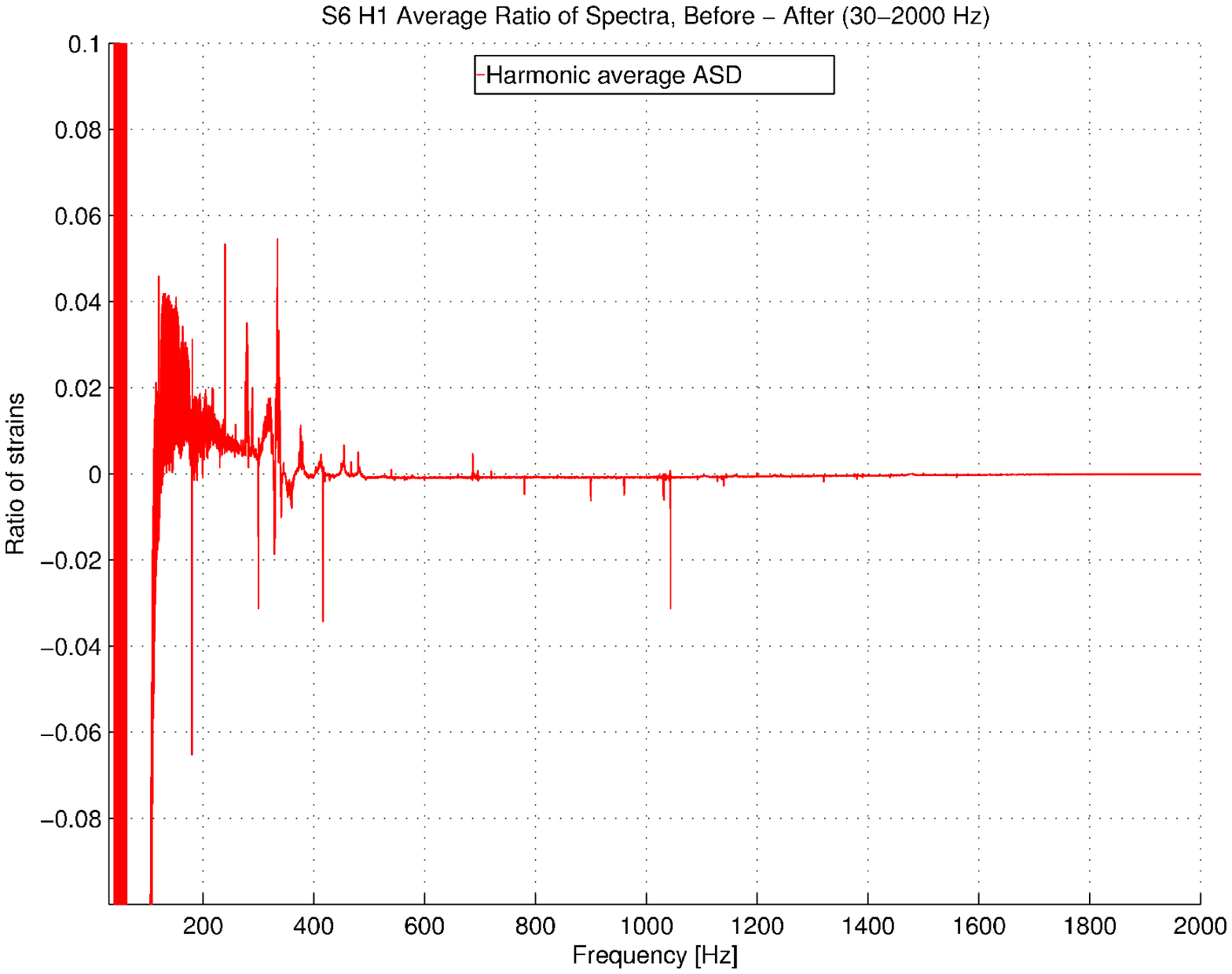}
\caption{Harmonic mean, GPS seconds $931.0 \times 10^6$ (2009 July 07) to $932.8\times 10^6$ (2009 July 28): \textit{(before-after)} (L), \textit{(before-after)/before} (R); greater than zero is improvement. The mean shows the absolute and relative difference of before and after, between the average of many spectra such as Figures~\ref{typicalInspiralGraph} and~\ref{bestInspiralGraph}, although the algorithm differs. Improvement from 80 to 400 Hz is noticeable; at higher frequencies there is degradation, negligible in relative terms, due to high-frequency filter rolloff. Frequencies below 50 Hz should be disregarded; they are usually not searched by LIGO, so spectra for this plot were generated with a high-pass filter at 38 Hz.}
\label{SFTgraph}
\end{center}
\end{figure}

Post-processing tests also compute average SFT spectra. For GPS second $931.0\times 10^6$ (2009 July 07; GPS seconds count from 1980 January 01) to $932.8\times 10^6$ (2009 July 28, about 10\% of S6), the harmonic mean spectra are shown in Figure~\ref{SFTgraph}.

SFTs are high-pass filtered at 38 Hz. The harmonic mean spectrum shows several percent improvement from about 80 Hz up to the 330 Hz violin mode frequencies. Above 400 Hz, there is proportionally-minor degradation,  due to filter rolloff. 

        \subsection{Feedforward benefits and potential}
        \label{benefits}

Inspiral range $\mathcal{R}$ increases for both S6 LIGO observatories, which should generalize to any observatory with broadband noise due to contamination from auxiliary servos.

\begin{figure}
\begin{center}
\includegraphics[height=75mm, width=150mm]{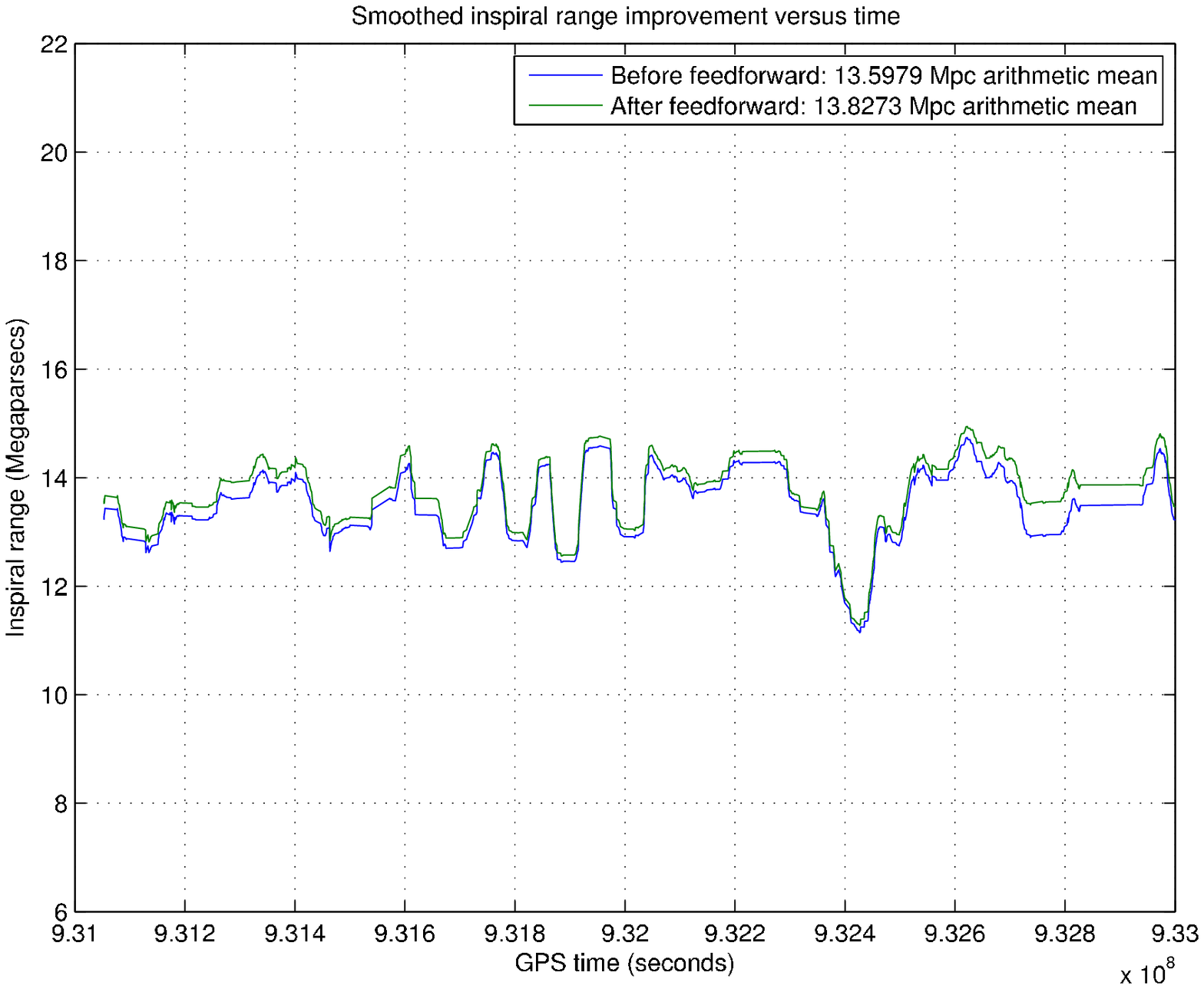}
\includegraphics[height=75mm, width=150mm]{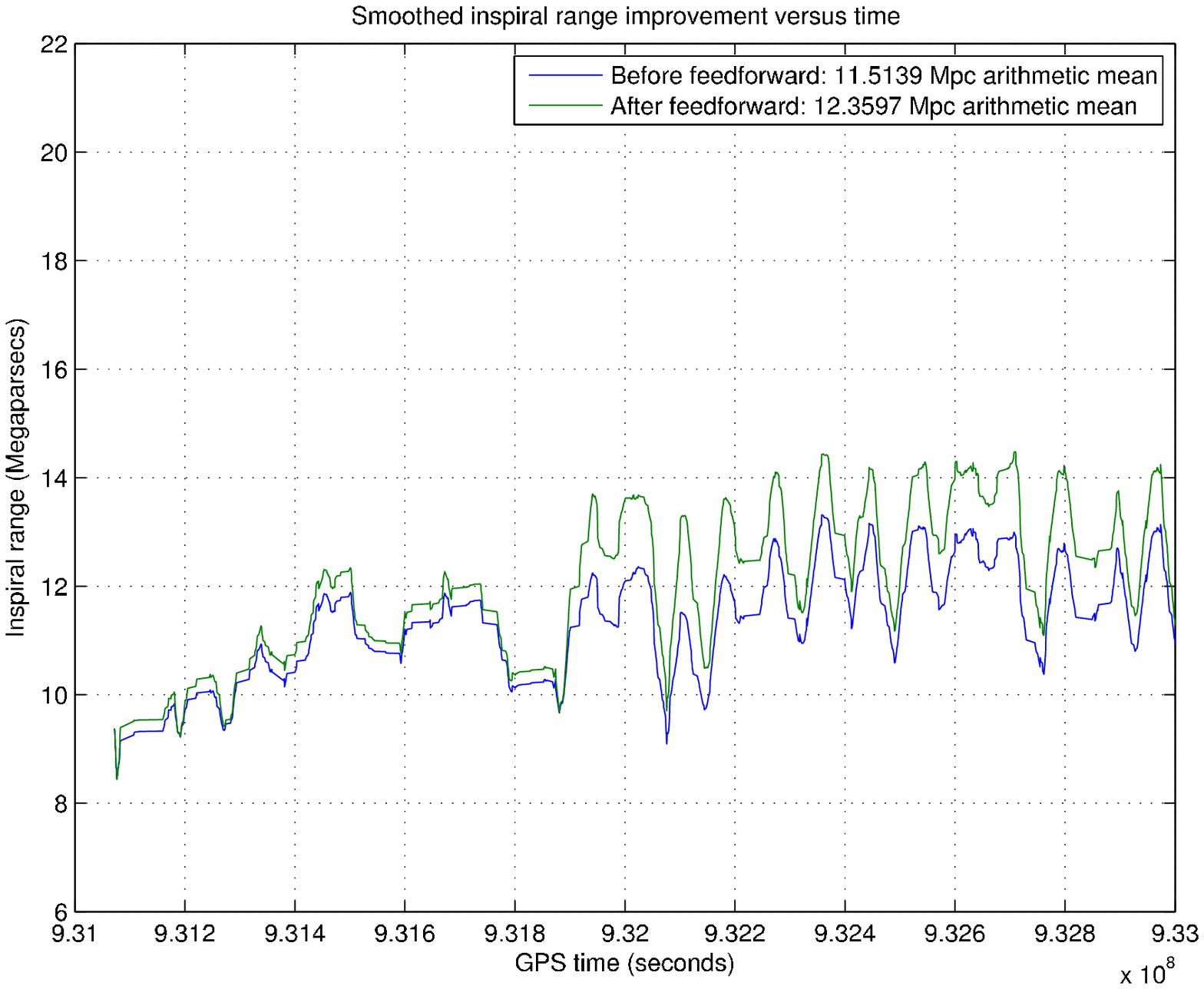}
\caption{Inspiral range vs time for S6 (starting 2009 July 07) before GPS time 9.33e8 (2009 July 30):
LIGO Hanford Observatory, H1 (top) gains 0.23 Mpc; LIGO Livingston Observatory, L1 (bottom) gains 0.84 Mpc. In this first month of S6, L1 saw greater benefit from post-facto feedforward correction; later data from H1 and L1 would improve by fluctuating amounts thanks to better real-time feedforward servos. Although H1 is less improved than L1 here, real-time tunings were made soon after.}
\label{S6inspiralRange}
\end{center}
\end{figure}
\begin{figure}
\begin{center}
\includegraphics[height=75mm, width=150mm]{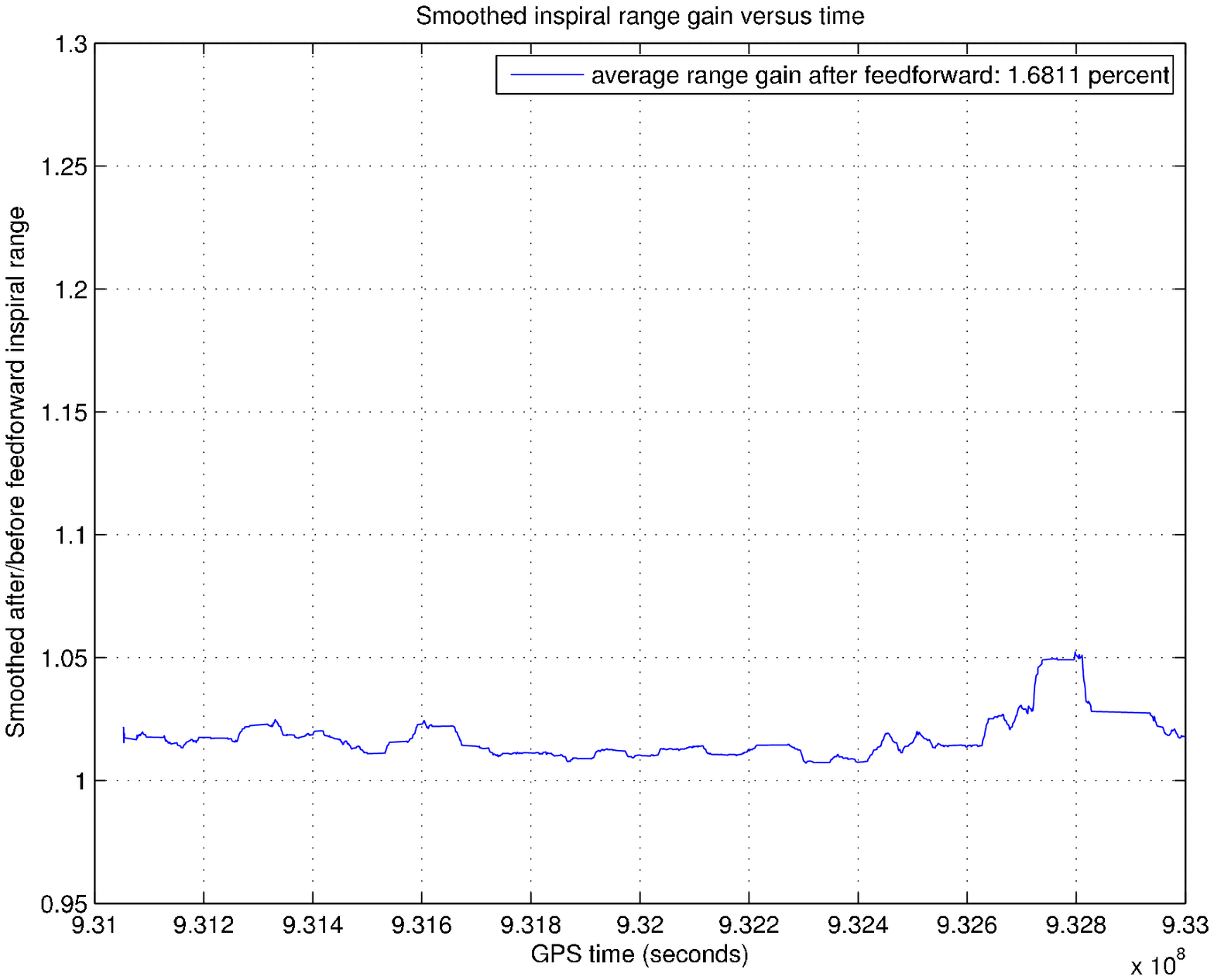}
\includegraphics[height=75mm, width=150mm]{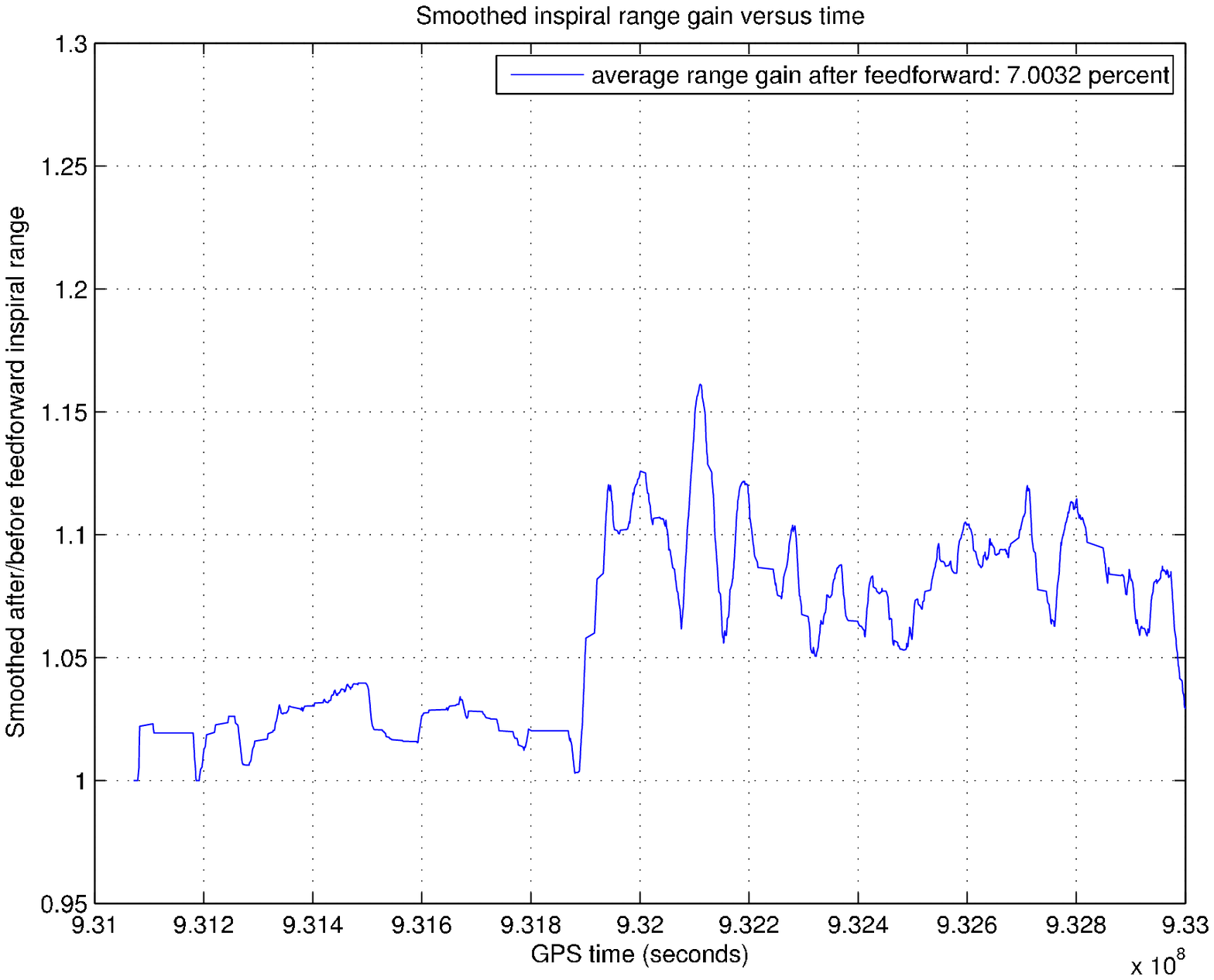}
\caption{Inspiral range \textit{fractional gain} vs time for S6 (starting 2009 July 07) before GPS time 9.33e8 (2009 July 30):
LIGO Hanford Observatory, H1 (top) 1.68\% better; LIGO Livingston Observatory, L1 (bottom) 7.00\% better. This plot shows relative gain for the same data for which Figure~\ref{S6inspiralRange} shows absolute gain.}
\label{S6inspiralRangeGain}
\end{center}
\end{figure}

Figures~\ref{S6inspiralRange} and~\ref{S6inspiralRangeGain} show the variation in achieved subtraction over about 10\% of S6.



\section{Conclusion}

Auxiliary MICH-PRC Subtraction has cleaned LIGO S6 data, yielding better strain sensitivity and inspiral range. Frequency-domain-derived, time-domain-applied feedforward correction removes noise by fitting a rational transfer function between witness \& target. Second order sections filter the witness channels, which then are subtracted from the measured target to produce an improved strain estimate. Diagnostics confirm that the corrected $h(t)$ benefits from dynamic, adaptive, algorithmic \textit{post facto} feedforward subtraction, gaining several percent in detectable inspiral range. Such an improvement potentially enhances the performance of any LIGO search. 

The subtraction leads to the lowest noise floor, around 150 Hz, of any time or interferometer so far (the highest performance to date at shot-noise limited frequencies has been obtained differently, with quantum optical squeezing~\cite{BarsottiNatureSqueezing,DwyerPhaseNoise}). This record may remain until Advanced LIGO. Thereafter, adaptive feedforward filters, real-time or \textit{post facto}, can be applied to mitigate noisy-but-inescapable couplings of the servo system. Signal recycling and filter cavities will further challenge commissioning. Angular and length sensing will need finer control servos. Advanced LIGO will also contain more physical and environmental monitors, from seismic and accelerometric to magnetic, that could provide witnesses for non-control-related noise. Altogether, more auxiliary channels and loops will exist, and while they may require sophisticated, non-linear methods, the subtraction technique presented here is a basis. Sensitive interferometry will benefit from simple, effective methods of suppressing instrumental influences.

\ack LIGO was constructed by the California Institute of Technology and Massachusetts Institute of Technology with funding from the National Science Foundation and operates under cooperative agreement PHY-0757058. This paper carries LIGO Document Number LIGO-P1300193. This research was also made possible by the generous support of the National Science Foundation, awards 0855422 and 1205173, LIGO Hanford Observatory, the LIGO Scientific Collaboration, and the University of Michigan. The authors wish to thank Gregory Mendell as well as Stuart Anderson, Juan Barayoga and Dan Kozak for grid computing expertise, Ian Harry for investigating signal recovery before and after injections and providing conclusions about signal-to-noise ratio for matched filtering, and Jeff Kissel for refining MICH and PRC subtraction by hand. Tobin Fricke wrote the converter function from second-order-system to ZPK filtering and also reviewed this manuscript, as did Rana Adhikari and Jenne Driggers, who developed many of these methods at the Caltech 40 m interferometer. Michael Coughlin, Jan Harms, and Nicol\'{a}s Smith-Lefebvre all generously provided comments. Finally, we thank the referees for Classical and Quantum Gravity for helpful suggestions.


\section*{References}
\bibliographystyle{jphysicsB}
\bibliography{Meadors_LIGO_AMPS_feedforward}

\end{document}